\documentclass{aa}  

\usepackage{graphicx}
\usepackage{txfonts}
\usepackage{color}
\usepackage{multirow}
\usepackage{soul}

\usepackage{subcaption}
\usepackage{xspace}
\usepackage{placeins}
\usepackage{hyperref}
\newcommand{\gaia}{{\it Gaia}\xspace}
\newcommand{\kms}{{$\rm{km}\,\rm{s}^{-1}$}\xspace}
\newcommand{\teff}{\ensuremath{T_{\mathrm{eff}}}\xspace}
\newcommand{\feh}{{[Fe/H]}\xspace}
\newcommand{\logg}{{$\log g$}\xspace}
\newcommand{\bprp}{{$\rm{BP}-\rm{RP}$}\xspace}
\newcommand{\ebv}{{E(B$-$V)}\xspace}

\begin{document} 

   \title{Radial velocities from Gaia BP/RP spectra}

   \author{Sill Verberne
          \inst{1}\fnmsep\thanks{Corresponding author: Sill Verberne\\
          \email{verberne@strw.leidenuniv.nl}}
          \and
          Sergey E.~Koposov\inst{2,3,4}
          \and
          Elena Maria Rossi\inst{1}
          \and
          Tommaso Marchetti\inst{5}
          \and
          Konrad Kuijken\inst{1}
          \and
          Zephyr Penoyre\inst{1}
          }

   \institute{Leiden Observatory, Leiden University,
              P.O. Box 9513, 2300 RA Leiden, the Netherlands
         \and
             Institute for Astronomy, University of Edinburgh, Royal Observatory, Blackford Hill, Edinburgh EH9 3HJ, UK
         \and
             Institute of Astronomy, University of Cambridge, Madingley Road, Cambridge CB3 0HA, UK
         \and
             Kavli Institute for Cosmology, University of Cambridge, Madingley Road, Cambridge CB3 0HA, UK
         \and
             European Southern Observatory, Karl-Schwarzschild-Strasse 2, 85748 Garching bei Munchen, Germany
             }

   \date{Received 2023 October 10; accepted 2024 January 15}
 
  \abstract
   {}
   {The \gaia mission has provided us full astrometric solutions for over $1.5$B sources. However, only the brightest 34M of those have radial velocity measurements. As a proof of concept, this paper aims to close that gap, by obtaining radial velocity estimates from the low-resolution BP/RP spectra that \gaia now provides. These spectra are currently published for about 220M sources, with this number increasing to the full $\sim 2$B \gaia sources with \gaia Data Release 4.}
   {To obtain the radial velocity measurements, we fit \gaia BP/RP spectra with models based on a grid of synthetic spectra, with which we obtain the posterior probability on the radial velocity for each object. Our measured velocities show systematic biases that depend mainly on colours and magnitudes of stars. We correct for these effects by using external catalogues of radial velocity measurements.} 
   {We present in this work a catalogue of about $6.4$M sources with our most reliable radial velocity measurements and uncertainties $<300$ \kms obtained from the BP/RP spectra. About 23\% of these have no previous radial velocity measurement in \gaia RVS. Furthermore, we provide an extended catalogue containing all 125M sources for which we were able to obtain radial velocity measurements. The latter catalogue, however, also contains a fraction of measurements for which the reported radial velocities and uncertainties are inaccurate.}
   {Although typical uncertainties in the catalogue are significantly higher compared to those obtained with precision spectroscopy instruments, the number of potential sources for which this method can be applied is orders of magnitude higher than any previous radial velocity catalogue. Further development of the analysis could therefore prove extremely valuable in our understanding of Galactic dynamics.}


   \maketitle
%

\section{Introduction}
The \gaia mission \citep{Gaia_2016} has been collecting data since 2014, with the primary scientific data products being the positions, proper motions and parallaxes of about $1.5$B objects. In the most recent data release \citep[DR3;][]{Gaia_2022}, low-resolution spectra have additionally been published of $\sim220$M objects. These spectra are obtained from two low-resolution prism spectrographs; BP observing in the wavelength range of 330--680 nm and RP in the 640--1050 nm range, collectively referred to as XP spectra. Their primary purpose is to provide source classification and astrophysical information on the astrometric sources observed; e.g.~stellar metallicity and line-of-sight extinction \citep{Bailer-Jones_2013}. In order to measure stellar parameters, such as radial velocities and elemental abundances, \gaia is equipped with the Radial Velocity Spectrometer \citep[RVS;][]{Katz_2023}. However, RVS spectra will not be available for all \gaia sources, being limited to $\mathrm{G}_{\mathrm{RVS}}\leq16$ in \gaia DR4 \citep{Katz_2023}. XP spectra on the other hand will be published in DR4 for all sources appearing in the astrometric catalogue with a limiting magnitude of $\rm{G}\approx20.7$ \citep{Gaia_2016}. The current magnitude limit of XP spectra in \gaia DR3 is $\mathrm{G}=17.65$ \citep{Gaia_2022}.\

XP spectra have been recognised as a rich source of astrophysical information, with efforts to measure [M/H], [$\alpha$/M] \feh, \logg, \teff, and line-of-sight extinction among others \citep[e.g.][]{Rix_2022, Zhang_2023, Andrae_2023_GSP, Andrae_2023, Guiglion_2023, Jiadong_2023}. This work focuses on obtaining radial velocity measurements for the first time from the low-resolution XP spectra. Although the precision of any radial velocity measurement from XP spectra is expected to be lower than that of conventional spectroscopic surveys, the scientific content would still be very significant due to the number of objects (220M currently and $\sim2$B in DR4). This would constitute a factor $\sim6.5$ increase in the total number of sources with radial velocity measurements compared to the currently largest radial velocity catalogue: \gaia DR3 with $\sim34$M measurements. Additionally, the recently launched Euclid space telescope \citep{Laureijs_2011} also includes a low-resolution slitless spectrograph. Although the Euclid instrument operates in the near infrared, the spectral resolution is significantly higher\footnote{\url{https://sci.esa.int/web/euclid/-/euclid-nisp-instrument}} compared to \gaia XP spectra, though still considered low-resolution. The method presented in this paper could in principle also be applicable to those data.\ 

This work should be seen as a proof of concept; focusing on demonstrating our ability to obtain radial velocity information from \gaia XP spectra, rather than obtaining the most accurate and precise measurements possible for all sub-types of objects appearing in \gaia DR3 XP spectra. We therefore advise readers to carefully consider if our measurements are appropriate for their specific use case.\

This paper is structured as follows: Section~\ref{sec:BP/RP spectra} discusses the properties and format of \gaia XP spectra, Section~\ref{sec:spectral_analysis} describes the analysis of the XP spectra to obtain radial velocity measurements, Section~\ref{sec:calibration} provides the post-calibration we apply to our radial velocity measurements using reference radial velocities, Section~\ref{sec:RFC} presents the random forest classifier we train for data quality assurance, Section~\ref{sec:Validation} provides validation of our calibrated results, Section~\ref{sec:Catalogue} describes our main and extended catalogues, which we publish together with this paper, in Section~\ref{sec:HVS} we discuss the science case of hypervelocity stars for our catalogue, in Section~\ref{sec:Discussion} we discuss our results and provide prospects for \gaia DR4, and lastly in Section~\ref{sec:Conclusion} we give closing remarks.

\section{BP/RP spectra}
\label{sec:BP/RP spectra}
In the following section we discuss a number of important points on the calibration and representation of XP spectra in \gaia DR3. We will only consider XP spectra from sources with $G<17.65$, which is the main XP catalogue from \gaia consisting of $\sim219$M sources. The few hundred thousand sources fainter than this limit mainly consist of white dwarfs and QSOs \citep{Gaia_2022}.\

\subsection{\gaia XP calibration}
\label{sec:gaia_xp_calibration}
The \gaia mission relies on self-calibration where possible. In the case of spectra, this is only possible to a limited degree. The calibration of multiple measurements, taken possibly years apart using different CCDs and fields-of-view, into a single mean spectrum is described by \citet{DeAngeli_2023} and is done using self-calibration. The calibration onto a physical wavelength and flux scale is described by \citet{Montegriffo_2023} and is performed using external measurements.

At wavelengths below 400 nm and above 900 nm, the wavelength calibration is less accurate, because there is an insufficient number of calibrator QSOs with emission lines in that part of the spectrum \citep[see Fig.~19][]{Montegriffo_2023}. In addition, there is a systematic offset in the RP spectra \citep{Montegriffo_2023}, which would lead to a systematic offset in radial velocity if not corrected for. The exact origin of this offset is unknown, but \citet{Montegriffo_2023} note that it might be caused by a systematic error in the line-spread-function model.

In terms of flux calibration, the uncertainties are typically underestimated \citep[see Fig.~18][]{DeAngeli_2023}. This underestimation is more pronounced for bright sources, but present in the majority of spectra and wavelength dependent. The underlying cause of this underestimation is unknown.\ 

\subsection{Basis function representation}
\label{sec:basis_function_representation}
\gaia observes the same sources multiple times over a time span of years, using two different fields-of-view and an array of CCDs \citep{Gaia_2016}. Small differences in the dispersion, wavelength coverage, instrument degradation, etc.~between observations gives the opportunity to extract more spectral information (i.e.~higher resolution spectra) from the sources than would be possible given a single observation. Representing this information in flux-wavelength space (henceforth referred to as sampled spectra), would be highly inefficient due to the small nature of the variations compared to the spectral resolution. For this reason, the \gaia consortium instead chose to represent the spectra as a series of coefficients for basis functions that describe the spectra. The BP and RP spectra are represented by 55 such spectral coefficients each, making for a total of 110 coefficients that describe every source. The first few coefficients contain most of the spectral information, since they are optimised for representing 'typical' \gaia sources \citep{DeAngeli_2023}. Alongside the spectral coefficients, \gaia has published their uncertainties and correlation coefficients, allowing us to construct the full covariance matrix for the coefficients of each source.

While providing more information than a sampled spectrum with the same number of samples, the representation in spectral coefficients also introduces challenges. Due to the individual basis-functions being continuous functions over the entire wavelength range, the uncertainties on all spectral coefficients are correlated. This means that when converting the basis function representation into sampled flux-wavelength space, all data points are correlated. Random noise in the initial \gaia observations in particular, causes random wiggles in the sampled XP spectra that could be mistaken for physical spectral features.\

\section{Spectral analysis}
\label{sec:spectral_analysis}
Now that we have discussed some of the important features of the XP spectra, we will first describe our spectral analysis of these data to obtain radial velocity measurements. 

\begin{figure*}
\centering
\begin{subfigure}{.48\textwidth}
\includegraphics[width=\linewidth]{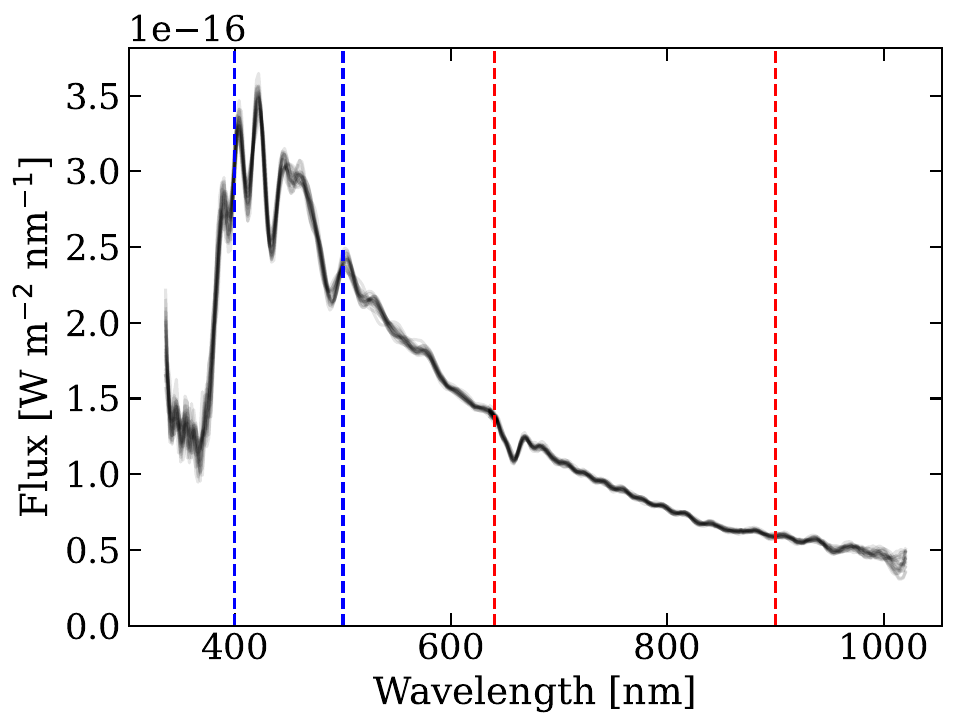}
\end{subfigure}
\begin{subfigure}{.48\textwidth}
\includegraphics[width=\linewidth]{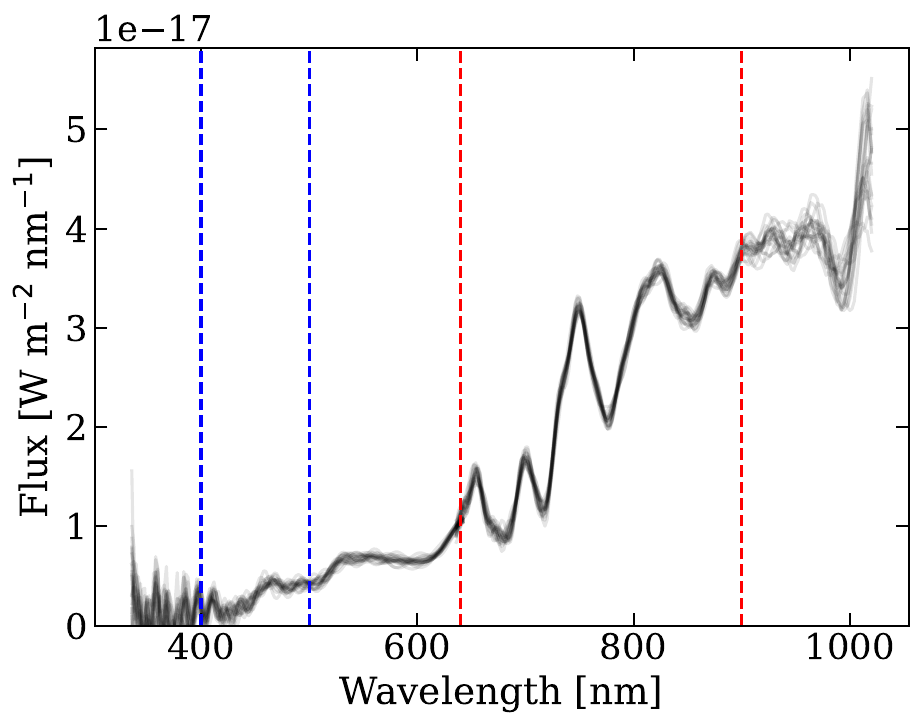}
\end{subfigure}
\caption{Example of two sampled \gaia XP spectra in black. The uncertainties in the spectral coefficients are sampled over to indicate the uncertainties in the sampled spectra. The blue and red dashed lines indicate the BP and RP spectral ranges used in the fitting procedure respectively. On the left a hot star is shown with \gaia DR3 \texttt{source\_id} \object{191594196746880} displaying prominent Balmer features, while on the right we show a red source with \gaia DR3 \texttt{source\_id} \object{31958852451968} containing broad molecular absorption bands.}
\label{fig:example fit}
\end{figure*}

In this paper we choose to convert the spectral coefficients to sampled spectra using \texttt{GaiaXPy}\footnote{\url{https://gaia-dpci.github.io/GaiaXPy-website/}\

\url{https://dx.doi.org/10.5281/zenodo.7566303}}. The conversion is performed through the design matrix as 
\begin{equation}
    \label{eq:coeff_to_samp}
    \bf{s} = \mathrm{A}\cdot\bf{b},
\end{equation}
with $\bf{s}$ the mean sampled spectrum, $\mathrm{A}$ the design matrix provided by \texttt{GaiaXPy}, and $\bf{b}$ the spectral coefficients. This gives us two spectra for each source, BP and RP, which we choose to sample on a grid of $\Delta\lambda=2$ nm. However, we do not use the entire spectra for our analysis: we select the wavelength range 400--500 nm for the BP spectra, while for the RP spectra we select 640--900 nm. Two example sampled \gaia XP spectra are shown in Fig.~\ref{fig:example fit}.
The BP-range is chosen to include the prominent Balmer lines, but exclude the region below 400 nm, where the wavelength calibration might be problematic, and the region above 500 nm, where for many stars the continuum would dominate the fit. The RP-range is much wider and includes most of the RP spectral range, except the region above 900 nm, where again the wavelength calibration might be problematic (see Sect.~\ref{sec:gaia_xp_calibration}).\\

Now that we have discussed how we handle the data, we describe in the following how we produce model spectra. We create the model $\mathrm{\bf{M}}(T_{\mathrm{eff}},\ \log g,\ [\mathrm{Fe/H}],\ v_\mathrm{r},\ \mathrm{E(B-V)})$, where \teff, \logg, and \feh correspond to the effective temperature, surface gravity, and metallicity from the PHOENIX spectral library respectively \citep{Husser_2013}, $v_\mathrm{r}$ is the radial velocity, and \ebv the extinction along the line-of-sight. For the PHOENIX models, we only consider atmospheres with $[\alpha/\rm{H}]=0$ for computational reasons. We shift each of these models by a radial velocity with a step size of 30 \kms. The parameter ranges for radial velocity, \teff, \logg, and \feh are displayed in Table~\ref{tab:grid edges}. 
\begin{table}[]
    \centering
    \caption{Extrema of the parameter ranges of the grid.}
    \begin{tabular}{lcc}
    \hline
         Parameter &  Minimum & Maximum \\
         \hline 
         Radial velocity (\kms) & $-3000$  & $3000$\\
         \teff (K) & 2300 & 15000\\
         \logg (dex) & $-0.5$ & 6.5\\
         \feh (dex) & $-3.0$ & $1.0$\\
    \hline
    \end{tabular}
    \label{tab:grid edges}
\end{table}
The resulting models are convolved with the resolution of the externally calibrated XP spectra. This is done by interpolating the values from Table 1 in \citet{Montegriffo_2023}. These interpolated values are then used at each wavelength in the sampled data to spread the flux in wavelength space using a Gaussian with standard deviation $\sigma = \mathrm{FWHM}/\left(2\sqrt{2\mathrm{ln}2}\right)$. Lastly, we apply extinction on a source-to-source basis using the 2D extinction map from \citet{Schlegel_1998} and the re-calibration from \citet{Schlafly_2011}, where we assume all sources to be behind the extinction layer. The extinction law we use is from \citet{Fitzpatrick_1999}. \\

Having described how we prepare the XP spectra and create model spectra, we will now describe how we fit the spectral models to the XP spectra. Given an XP spectrum from \gaia, we can determine the likelihood of the data given a model, $P(\mathbf{D}\mid \mathbf{M})$, from
\begin{equation}
P(\mathbf{D}\mid\mathbf{M}) = \frac{1}{\sqrt{(2\pi)^k|\mathbf{C}|}}\exp\left(-\frac{1}{2} \left[\mathbf{D} - \mathbf{M}\right]^{\top} \mathbf{C}^{-1} \left[\mathbf{D} - \mathbf{M}\right]\right),
\end{equation}
where $\bf{D}$ is the data, $\bf{M}$ the model, $k$ the number of dimensions, and $\bf{C}$ the covariance matrix of the data \citep[e.g.][]{Hogg_2010}. This holds in the case where the uncertainties are Gaussian with correctly estimated variances, which is not strictly true in our case. Because we over-sample our sampled spectra with respect to the orthogonal bases, our covariance matrix in sampled space does not have full rank. To allow inversion of the covariance matrix in sampled space, we only consider the diagonal elements and thus discard correlation information. This causes the uncertainties on the radial velocity measurements to be further underestimated.

We calculate this likelihood for all models, after which we marginalise over the nuisance parameters T$_{\rm{eff}}$, \logg, and \feh. We use flat priors on our parameters between the extrema of the parameter grid shown in Table~\ref{tab:grid edges}. \teff is the exception; for computational reasons we instead only consider models differing no more than 500K from an initial guess on \teff we make based on the \bprp colour of a source and the extinction. The initial guess on \teff is described in Appendix \ref{app:teff}. During analysis of the results, we noticed that the performance of this initial guess is poor for \ebv $\gtrsim0.5$. For this reason we only report results for sources with \ebv $<0.5$, for which the method works well.

In order to determine the radial velocity and corresponding uncertainty we assume a Gaussian posterior probability on the radial velocity. We fit a parabola to the log-posterior probability by selecting all radial velocity points with a log-posterior probability no less than 10 from the maximum log-posterior probability. If less than 5 points in radial velocity space meet this requirement, we reduce the threshold by increments of 10 until we have more than 5 points. If the resulting fit peaks outside our radial velocity range of $\pm3000$ \kms, we consider the fit to have failed and report no radial velocity. \\

Now that we have laid down the foundation of our method we will describe the skewness and goodness-of-fit measurements we use to evaluate the reliability of the radial velocity (uncertainty) measurements in Sect.~\ref{sec:RFC}. 

By fitting a parabola to the log-posterior probability, we are assuming symmetric uncertainties. To evaluate if this is a reasonable assumption, we determine the skewness for the log-posterior probability distribution using
\begin{equation}
g_1 = \frac{\sum_{i=1}^{n} P_i \left({v_\mathrm{r}}_i - \overline{v_\mathrm{r}}\right)^3}{\left(\sum_{i=1}^{n} P_i \left({v_\mathrm{r}}_i - \overline{v_\mathrm{r}}\right)^2\right)^{3/2}},
\label{eq:skew}
\end{equation}
where $P_i$ is the posterior probability per radial velocity bin, ${v_\mathrm{r}}_i$ the corresponding radial velocity, and $\overline{v_\mathrm{r}}$ the mean radial velocity given by
\begin{equation}
\overline{v_\mathrm{r}} = \sum_{i=1}^{n} {v_\mathrm{r}}_i\cdot P_i.
\end{equation}
This allows us to identify cases in which the posterior probability distribution is asymmetric and for which the symmetric uncertainties might not be reliable. In addition, we calculate the reduced $\chi^2$ of our best fit, by approximating the number of degrees of freedom as the number of data points we have (i.e. number of flux vs wavelength points) minus the number of parameters we fit (4).

\subsection{Results of spectral analysis}
\label{sec:results_analysis}
Here we will discuss the results from the spectral analysis presented above. As mentioned, the analysis was applied to all $\sim219$M XP sources with $G<17.65$. There are generally three outcomes possible for our spectral analysis. The first outcome is that we obtain a measurement for the radial velocity and corresponding uncertainty of a particular source. It is also possible that a fit failed, because the best fit radial velocity was outside our parameter range of $\pm3000$ \kms, or a column  in \gaia, such as the BP colour, required by our processing was not measured or was unavailable. The third outcome is that the initial guess for \teff was outside our model range, in which case we do not perform a fit (see Table~\ref{tab:grid edges}). We summarise the relevant numbers in Table~\ref{tab:summary}.
\begin{table}[]
    \centering
    \caption{Summary of the raw results of this work, indicating the number of sources analysed and the number of those for which we obtained a radial velocity estimate.}
    \begin{tabular}{lc}
    \hline
         XP spectra analysed &  218\,969\,408\\
         Predicted \teff\ outside grid range & 15\,041\,788\\
         \ebv $\geq 0.5$ & 55\,594\,623\\
         Failed fits & 23\,187\,507\\
         \hline
         Radial velocities obtained & 125\,145\,490\\
    \hline
    \end{tabular}
    \label{tab:summary}
\end{table}

\section{Radial velocity calibration}
\label{sec:calibration}
Because of the calibration issues described in Sect.~\ref{sec:BP/RP spectra} we expect to see systematic offsets in our measurements of radial velocities that are a function of colour, magnitude, and extinction, in addition to an underestimation of uncertainties. To make matters more complicated, we expect the presence of an "outlier" population, which is a population of objects for which the measured radial velocity spread is well beyond formal errors. In this section we describe how we correct for these systematics in our radial velocities and their uncertainties obtained from the spectral analysis described in the previous section. We make use of reference radial velocity measurements from dedicated radial velocity surveys. We begin by describing our set of reference radial velocities in Sect.~\ref{sec:reference_data_set}, followed by the statistical model to describe our XP radial velocities compared to the reference measurements in Sect.~\ref{sec:calibration_models}, and finally the fitting procedure of the model to the data in Sect.~\ref{sec:fitting}.

\subsection{Reference data set}
\label{sec:reference_data_set}
The reference radial velocity measurements we use are \gaia RVS DR3 \citep{Katz_2023}, LAMOST DR8 low-resolution \citep{LAMOST_2012}, and APOGEE DR17 \citep{APOGEE_2017, APOGEEdr17}. These catalogues were chosen because of their large size and sky coverage. Importantly, APOGEE and LAMOST contain sources fainter than the \gaia RVS magnitude cut, which is needed to calibrate and validate our results for faint sources. A summary of the relevant statistics from these catalogues is included in Table~\ref{tab:external}.
\begin{table}[]
    \centering
    \caption{Summary of the reference radial velocity catalogue used to calibrate and validate our results. The combined size is smaller than the sum, since there is overlap between the surveys.}
    \begin{tabular}{lcc}
    \hline
         Catalogue & Radial velocities &  Magnitude limit \\
         &matched to BP/RP&\\
         \hline 
         \gaia (RVS) & 22\,018\,897 & G$_{\rm{RVS}} \leq14\tablefootmark{a}$ \\
         LAMOST & 3\,666\,919 & $\rm{r}\leq17.8$\tablefootmark{b}\\
         APOGEE & 445\,137 & $\rm{H}\lesssim12.8$\tablefootmark{c}\\
         \hline 
         Combined & 23\,900\,765 & \\
    \hline
    \end{tabular}
    \tablefoot{
    \tablefoottext{a}{\citet{Katz_2023}};
    \tablefoottext{b}{\citet{Yan_2022}};
    \tablefoottext{c}{\citet{Santana_2021}}}
    \label{tab:external}
\end{table}
The \gaia RVS radial velocities are measured with a different instrument and technique and can therefore be considered fully independent of the measurements we provide here. Cross-referencing LAMOST and APOGEE was done using the \gaia\ DR3 \texttt{source\_id} provided for each measurement by both LAMOST and APOGEE. When more than a single measurement was available for either LAMOST or APOGEE, we took the median of all measurements. This provides us with a total number of 23\,900\,765 sources for which we have both an XP and reference radial velocity measurement. We consider reference radial velocity measurements to be 'ground truth' and do not consider uncertainties in them. The reason is that our measurements will have uncertainties much larger than typical uncertainties in any of the reference catalogues. In our calibration we only consider sources that have no neighbours in \gaia within 2 arcseconds. The reason for this is that these sources tend to have blended spectra, due to the size of the spectral extraction window for XP spectra. Our models are not set up to account for blending, which means that radial velocity uncertainty and offset will be different for many of these sources. This further reduces the total number of sources used in calibration to 22\,397\,143.\

\subsection{Calibration model}
\label{sec:calibration_models}
To characterise the systematics in our radial velocity measurements we adopt a Gaussian mixture model with a likelihood given by
\begin{equation}
\label{eq:likelihood}
\begin{split}
\mathcal{L} \; \propto \; & \prod_{i=1}^{N}\left[\frac{1-f}{\sqrt{2\pi\sigma_i^2}}\:\exp\left(-\frac{\left[{v_\mathrm{r}}_\mathrm{ref} - {v_\mathrm{r}}_\mathrm{xp} - b\right]^2}{2\sigma_i^2}\right) \right. \\
&\left.+\: \frac{f}{\sqrt{2\pi\sigma^2_\mathrm{out}}}\:\exp\left(-\frac{\left[{v_\mathrm{r}}_\mathrm{xp}-y\right]^2}{2\sigma^2_\mathrm{out}} \right) \right],
\end{split}
\end{equation}
with $f$ the outlier fraction, $\sigma_i$ the radial velocity uncertainty, ${v_\mathrm{r}}_\mathrm{ref}$ the reference radial velocity measurement, ${v_\mathrm{r}}_\mathrm{xp}$ the XP radial velocity, $b$ the systematic offset between the reference and XP radial velocities, $\sigma_\mathrm{out}$ the standard deviation of the outlier population, and $y$ the offset of the outlier population. The uncertainties on radial velocity measurements ($\sigma_i$) are described by 
\begin{equation}
\label{eq:corr uncertainty}
\sigma_i = \sqrt{(a\sigma_\mathrm{m})^2 + c^2},
\end{equation}
with $a$ the underestimation factor on the uncertainties, $\sigma_\mathrm{m}$ the uncertainty determined from the posterior probability, and $c$ the noise floor parameter.

We use bins in \bprp colour, apparent G magnitude, and extinction to fit for the free parameters $f$, $a$, $b$, $c$, $\sigma_\mathrm{out}$, and $y$. We use 20 equally spaced bins in the range $5\leq G \leq 17.65$ and 40 bins in the range $-0.3\leq$ \bprp $\leq5$. For extinction we use bins with a width of $\Delta$\ebv$=0.1$. Additionally, we use two bins in \logg for sources with \bprp $\geq1.6875$, with a divide at \logg$=3.5$. We use the \logg measurements from \citet{Zhang_2023} for this purpose. The split in \logg is used because we observe a high degree of systematic offset in the radial velocities we measure between dwarfs and giants at these colours. A description and justification for this split in \logg is given in Appendix~\ref{app:logg_systematics}. We require at least 64 sources in a particular bin for fitting, with a maximum of 100\,000, above which we select 100\,000 sources from the sample at random for computational efficiency. We run the same calibration procedure using 10 equally spaced bins in the range $5\leq G \leq 17.65$ and 20 bins in the range $-0.3\leq$ \bprp $\leq5$, i.e.~using bins twice the default size. This makes sure that we have calibration for most sources, even in sparsely populated areas of the colour-magnitude space. If there are still not enough sources in the colour-magnitude bin for a particular source, we do not apply calibration.

\subsection{Fitting of the model}
\label{sec:fitting}
To estimate the parameters in our calibration model we use the Markov Chain Monte Carlo (MCMC) implementation in \texttt{emcee} \citep{Foreman_2013}. We use flat priors throughout, except for $\sigma_{\rm{out}}$, for which we use a log-uniform prior. Our MCMC approach is as follows: we initialise 64 walkers that we first propagate for 1000 steps to explore the parameter space. To avoid walkers getting stuck in local minima, we reject walkers that finished with a log-likelihood outside $8.4$ of the maximum log-likelihood over all walkers. The value 8.4 makes sure that 99\% of the walkers would remain if they traced a 6D Gaussian distribution. A next 1000 steps are performed with again 64 walkers that have been drawn randomly from the last 100 steps of the walkers that remained from the previous run. The last 800 steps of this run are used to compute the medians of the free parameters.

\subsection{Results of radial velocity calibration}
\label{sec:results_calibration}
The number of calibrated radial velocities is 123\,835\,034 out of a total of 125\,145\,490. This means that calibration was performed for $\sim99$\% of our radial velocity measurements from Sect.~\ref{sec:spectral_analysis}.

\begin{figure}
\centering
\includegraphics[width=\linewidth]{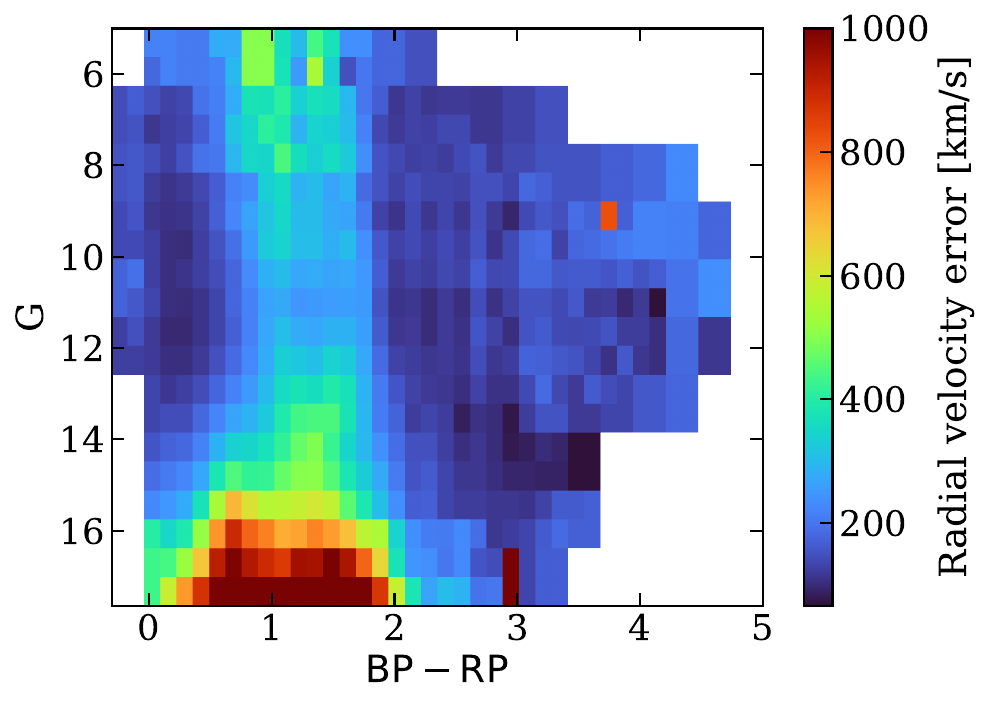}
\caption{Median calibrated uncertainties as a function of colour and apparent magnitude for sources with \ebv$<0.1$.}
\label{fig:uncertainties}
\end{figure}

Here we present the radial velocity calibration results for low extinction (\ebv$<0.1$) sources. Results for higher extinction sources are similar, unless specified. We show the calibrated uncertainties (see Eq.~\ref{eq:corr uncertainty}) on the radial velocities measured from the XP spectra in Fig.~\ref{fig:uncertainties}.
In the region $\rm{BP}-\rm{RP}\geq1.6875$, where we use two bins in \logg, we take the source-number average for each bin between the giants and dwarfs. In general we can see that the lowest uncertainties are obtained from blue and red sources. We observe higher uncertainties for $1\lesssim$ \bprp $\lesssim2$ and the uncertainty generally increases for faint sources. In addition, the figure shows that uncertainties down to $\sim100$ \kms are possible for red and blue sources. The reason why uncertainties are relatively high for $1\lesssim$ \bprp $\lesssim2$ is that there are few spectral features in the XP spectra for those sources. Without strong spectral features such as the Balmer lines and molecular absorption bands (see Fig.~\ref{fig:example fit}), fitting for a radial velocity becomes less precise.

\begin{figure}
\centering
\includegraphics[width=\linewidth]{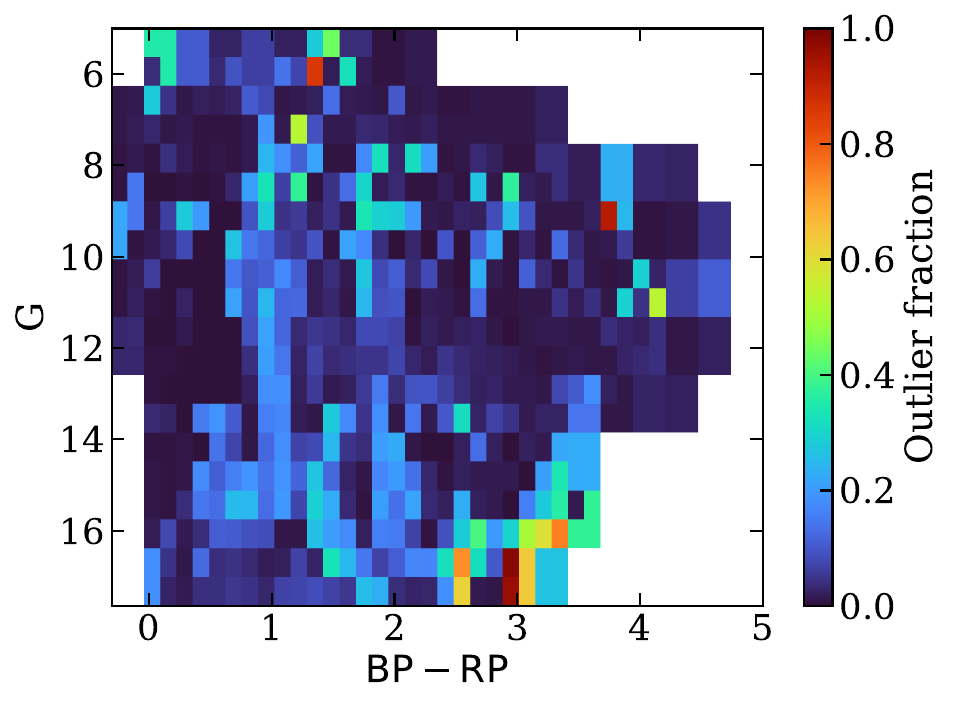}
\caption{2D histogram of the outlier fraction as a function of colour and apparent magnitude for sources with \ebv$<0.1$.}
\label{fig:outliers}
\end{figure}

In Fig.~\ref{fig:outliers} we show the outlier fraction as a function of colour and magnitude. 
The outlier fraction tends to be low (smaller than 0.1), with a few regions containing notably more outliers. For higher extinction, the outlier fraction increases substantially, which we show in Fig.~\ref{fig:outliers_app} in the Appendix. In the same Appendix we also show the underestimation factor, offset, and noise floor for sources with \ebv$<0.1$.

\begin{figure}
\includegraphics[width=\linewidth]{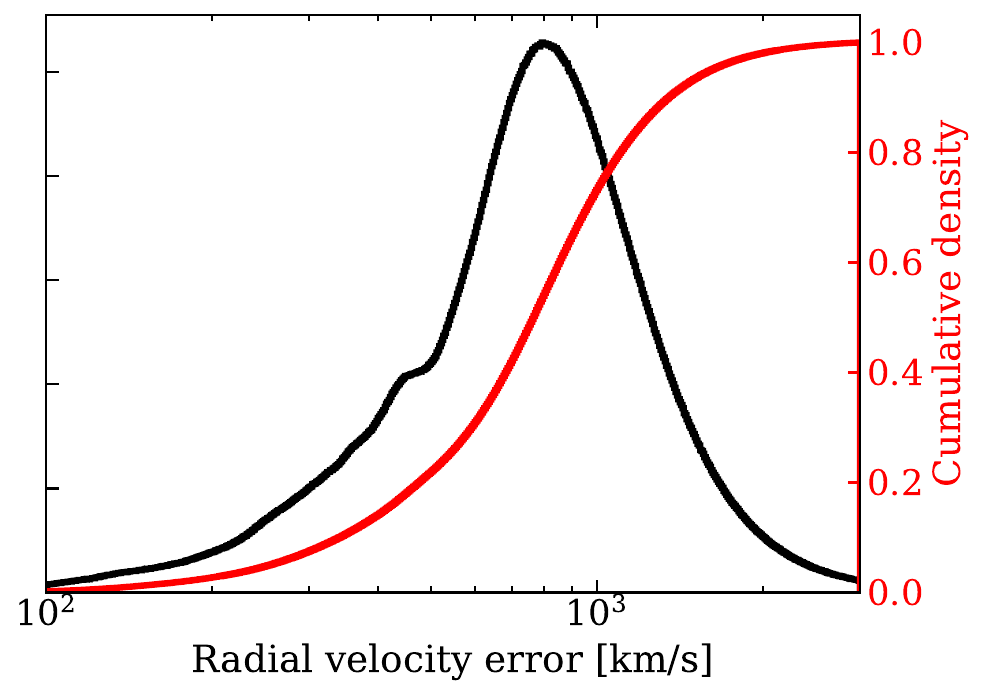}
\caption{In black we show the histogram of the calibrated radial velocity uncertainties. The solid red line shows the cumulative distribution.}
\label{fig:hist_uncertainties}
\end{figure}

Calibration has been applied to all sources when available. A histogram of the calibrated radial velocity uncertainties along with their cummulative distribution is included in Fig.~\ref{fig:hist_uncertainties}.
The median uncertainty on our calibrated radial velocity measurements is about $772$ \kms, but extends all the way down to below 100 \kms. Sources with low radial velocity uncertainty tend to be either blue (\bprp $\lesssim0.7$) or red (\bprp $\gtrsim2$).


\section{Random forest classifier}
\label{sec:RFC}
Although we have a general indication of the reliability of individual measurements from the outlier fraction parameter determined from our calibration model, a quality parameter determined on a source-to-source basis is important to avoid unreliable measurements in the final catalogue. We do this making use of a Random Forest Classifier (RFC). 

The definition for a bad measurement we use is $\Delta\mathrm{v_r}/\sigma_{\mathrm{vr}}>3$, with $\Delta\mathrm{v_r}$ being the difference between our calibrated measurement and the reference measurement and $\sigma_{\mathrm{vr}}$ being the corresponding calibrated uncertainty. For sources where we have access to reference radial velocities, we make use of 10-fold cross validation to predict the bad measurement probability. This makes sure the source for which we predict the bad measurement probability is never part of the training set. For the remaining sources, we train the RFC on all sources with reference radial velocity measurements. We take care to avoid information leaking from the training parameters to the radial velocities by excluding parameters like the sky coordinates and absorption. We use the \texttt{scikit-learn} RFC with 100 estimators \citep{scikit-learn} and the following parameters for training:
\begin{itemize}
    \item Reduced $\chi^2$ of our best fit model
    \item \teff, \logg, and \feh of the best fit model for radial velocity
    \item Extinction corrected \bprp colour of the source
    \item Skewness of the radial velocity posterior (see Eq.~\ref{eq:skew})
\end{itemize}
In addition to the following columns provided by the \gaia archive (see the \gaia documentation\footnote{\url{https://gea.esac.esa.int/archive/documentation/GDR3/Gaia_archive/chap_datamodel/sec_dm_main_source_catalogue/ssec_dm_gaia_source.html}\ 

\url{https://gea.esac.esa.int/archive/documentation/GDR3/Gaia_archive/chap_datamodel/sec_dm_spectroscopic_tables/ssec_dm_xp_summary.html\#xp_summary-}} for column descriptions):
\begin{itemize}
    \item \texttt{phot\_g\_mean\_mag}
    \item \texttt{phot\_xp\_mean\_mag}
    \item \texttt{xp\_n\_transits}
    \item \texttt{xp\_n\_blended\_transits}
    \item \texttt{xp\_n\_contaminated\_transits}
    \item \texttt{xp\_n\_measurements}
    \item \texttt{xp\_standard\_deviation}
    \item \texttt{xp\_chi\_squared/xp\_degrees\_of\_freedom},
\end{itemize}
where "xp" denotes that we use the corresponding column of both BP and RP. We found that the extinction corrected colour is the most important out of these, with a feature importance of 0.13. Most of the other columns have similar importance (between about 0.4 and 0.8), except for the blended and contaminated transits, which have low importance at $\lesssim0.2$. This procedure provides us with the likelihood that a particular measurement is unreliable, which we will refer to as the \texttt{bad\_measurement} parameter.

To verify the effectiveness of the RF, we looked at the outlier fraction as a function of this \texttt{bad\_measurement} parameter. We find good agreement with a one-to-one relation between the two, indication a successful classification.

\section{Validation}
\label{sec:Validation}
We have already discussed the results from our spectral analysis and calibration in Sects.~\ref{sec:results_analysis} and \ref{sec:results_calibration}. In addition, we have access to the quality parameter \texttt{bad\_measurement} described in Sect.~\ref{sec:RFC}. Using these earlier results, we focus in this section on validating that our measurements indeed measure the radial velocity and evaluate the reliability of our reported uncertainties.\\

\begin{figure}
\centering
\includegraphics[width=\linewidth]{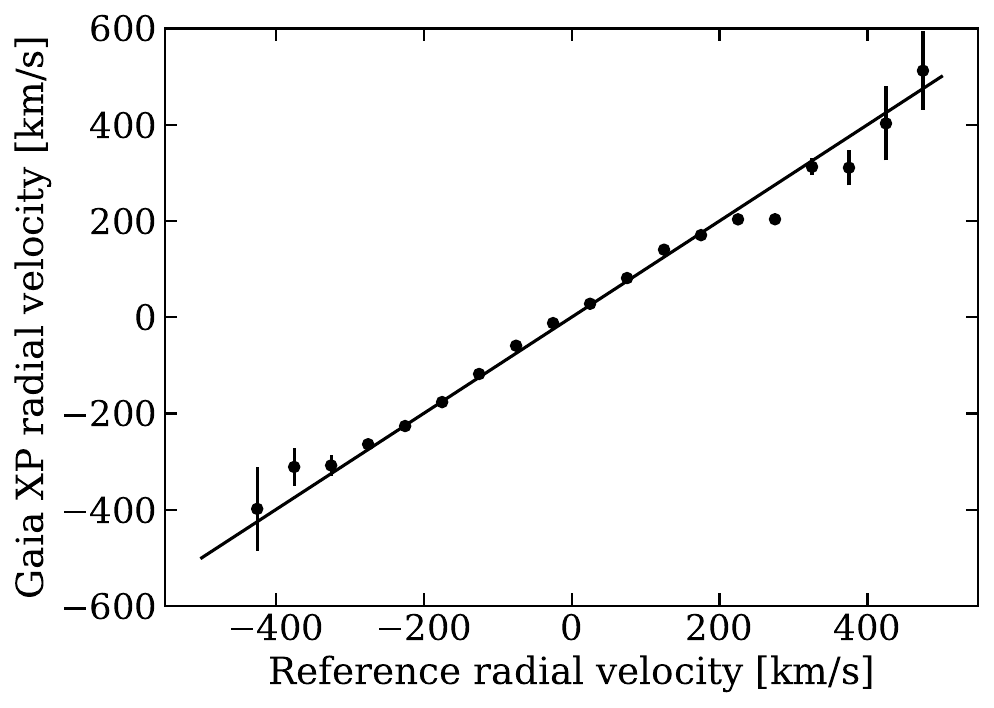}
\caption{Median of the calibrated binned XP radial velocities as a function of the reference radial velocity measurements. These sources have an outlier fraction below 0.2, \texttt{bad\_measurement} $<0.1$, and calibrated uncertainty below 300 km/s. The solid line is the bisection.}
\label{fig:binned_rv}
\end{figure}

\begin{figure*}
\centering
\begin{subfigure}{.44\textwidth}
\includegraphics[width=\linewidth]{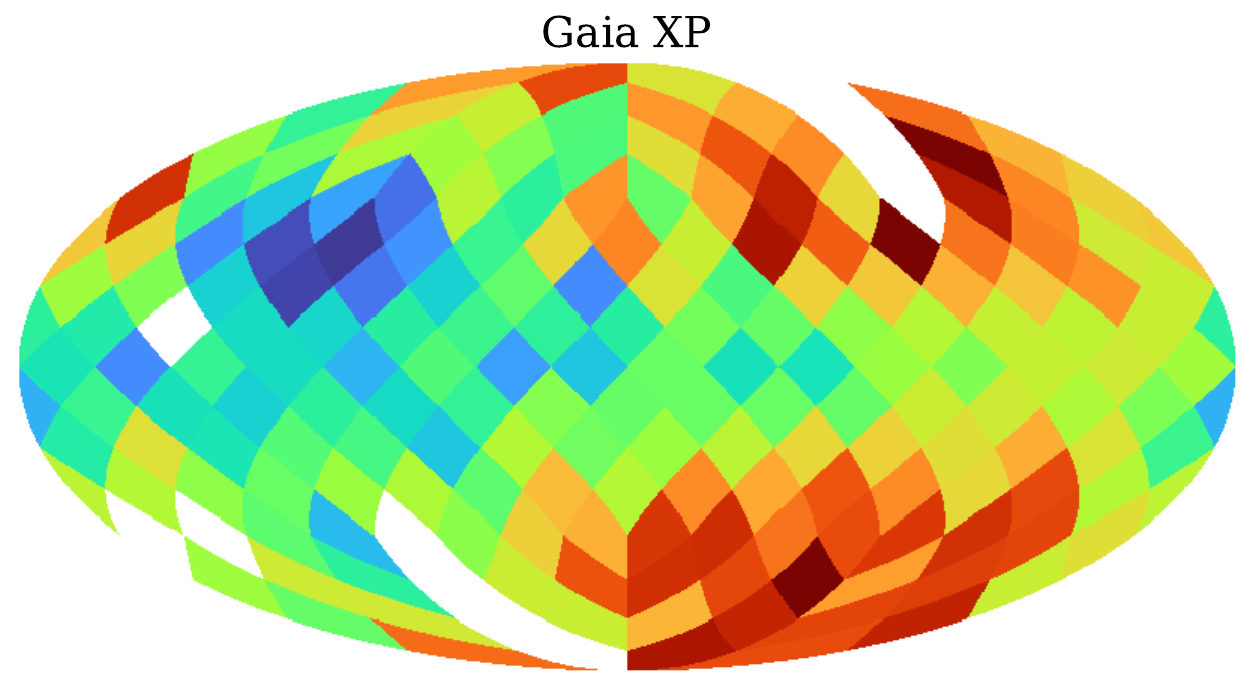}
\end{subfigure}
\begin{subfigure}{.44\textwidth}
\includegraphics[width=\linewidth]{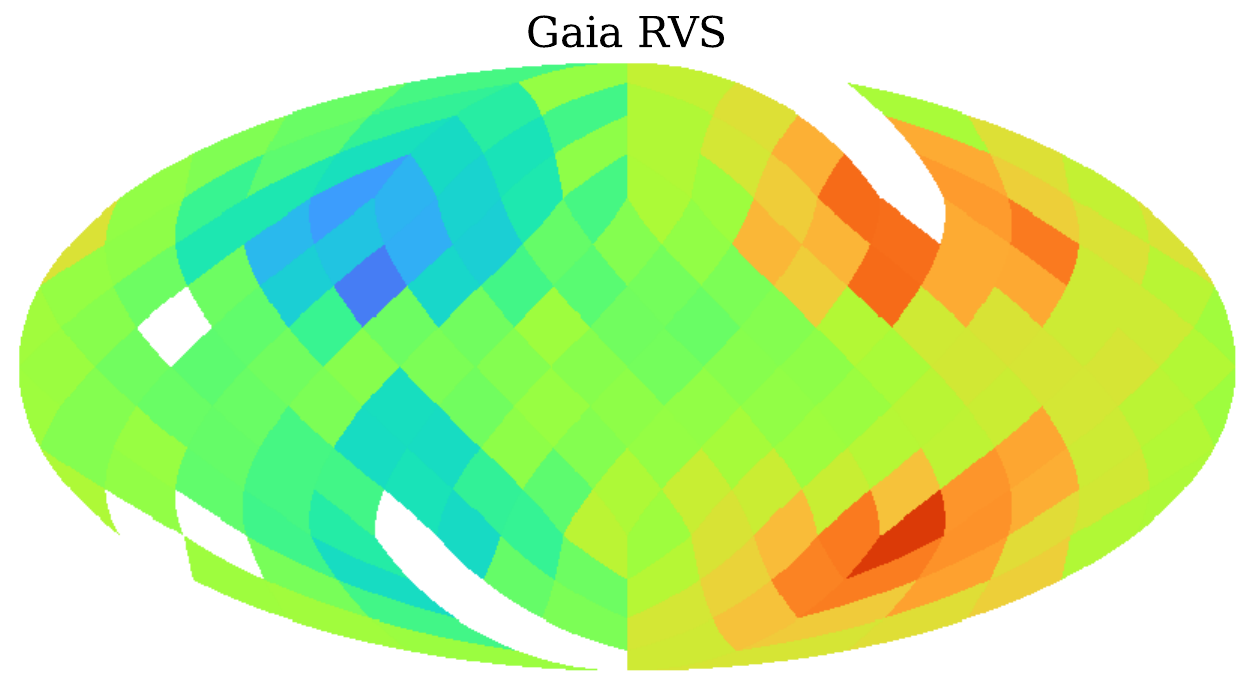}
\end{subfigure}
\begin{subfigure}{.07\textwidth}
\includegraphics[width=\textwidth]{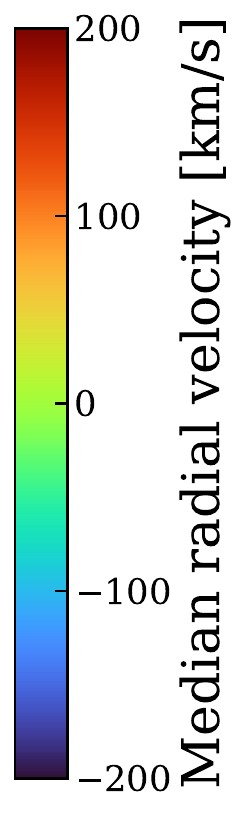}
\end{subfigure}
\caption{On the left we plot the median XP radial velocity as a function of sky position in Galactic coordinates and on the right the same thing using \gaia RVS reference measurements. We only look at low metallicity stars here (\feh$\leq-1$) to highlight halo stars. This increases the radial velocity amplitude as a function of position on the sky. What we can recognise is the dipole caused by the Solar motion in both maps. The other selections of the sources in this figure are described in the main text. The sources included in both panels are the same and we only colour bins with at least 10 measurements.}
\label{fig:sky_projection}
\end{figure*}

To demonstrate that we are indeed measuring radial velocities from the XP spectra, we include Fig.~\ref{fig:binned_rv}, in which we bin the XP radial velocities based on their reference measurements.
The uncertainties on the individual bins are calculated as
\begin{equation}
\sigma = \frac{1}{N}\sqrt{\frac{\pi}{2}} \left(\,\overline{|v_\mathrm{r} - \overline{v_\mathrm{r}}|^2}\, \right)^{1/2},
\end{equation}
where the overline indicates that the mean is taken, $v_\mathrm{r}$ is the radial velocity, and $N$ is the number of measurements in the bin. The measurements clearly follow the bisection with the reference radial velocity measurements, demonstrating that we indeed measure stellar radial velocities. To ensure we are not seeing the result of correlation of radial velocity and position in the colour-magnitude diagram picked up by our calibration model, we perform this analysis also for each colour-magnitude bin separately in Fig.~\ref{fig:veri_rv_sens} in the appendix. To further demonstrate our ability to constrain radial velocities, we plot in Fig.~\ref{fig:sky_projection} the sky projection of the median XP radial velocities compared to \gaia RVS.
The quality cuts we apply to our catalogue are 
\begin{itemize}
    \item \texttt{rv\_err} $<300$ \kms
    \item \texttt{CMD\_outlier\_fraction} $<0.2$
    \item \texttt{bad\_measurement} $<0.1$.
\end{itemize}
In addition, we use the catalogue from \citet{Zhang_2023} and select sources with \feh$\leq-1$ to mainly select halo stars and \texttt{quality\_flags}$\leq8$. This makes us able to see the dipole caused by the Solar motion in both the XP and RVS maps. The dipole disappears at the Galactic plane due to the sample being dominated by non-halo stars in that region.

\begin{figure}
\centering
\includegraphics[width=\linewidth]{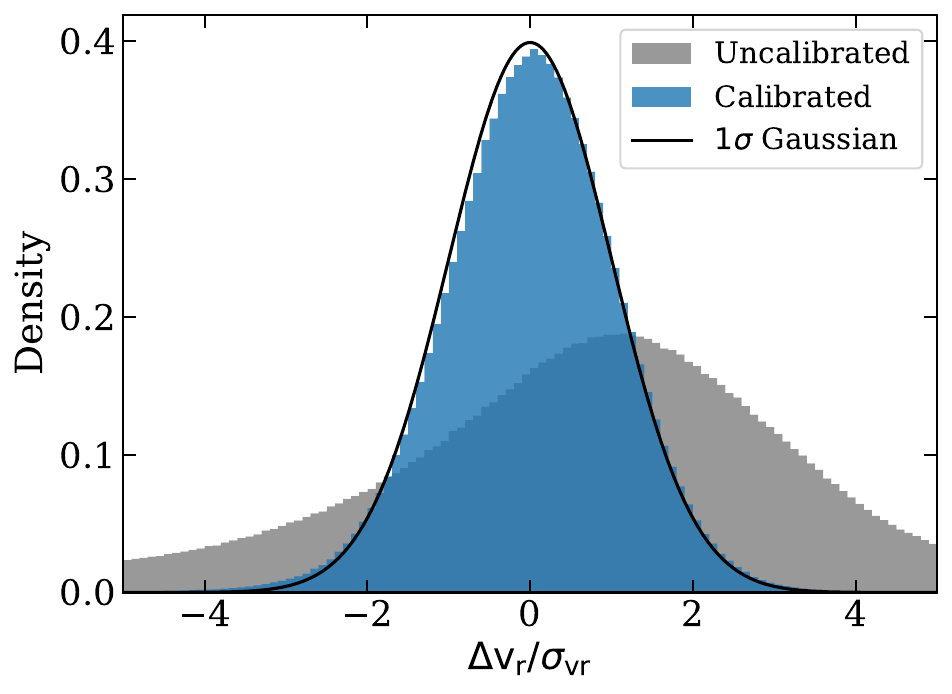}
\caption{The difference in radial velocity over the uncertainty of both our calibrated and uncalibrated results compared to reference measurements. The solid line is a Gaussian distribution with a standard deviation of 1. The sample used to make this figure has calibrated radial velocity uncertainties $<300$ \kms.}
\label{fig:exp_std}
\end{figure}

To evaluate if our radial velocity uncertainties are accurate, we provide the histogram of the radial velocity difference of our measurements compared to the reference measurements over the uncertainty in Fig.~\ref{fig:exp_std}.
We can determine the standard deviation of this distribution by
\begin{equation}
    \sigma = 1.4826\cdot \mathrm{median}\left(\Delta v_r/\sigma_{vr} - \widetilde{\Delta v_r/\sigma_{vr}} \right),
\end{equation}
with $\Delta v_r/\sigma_{vr}$ the difference between our radial velocity and the reference one over the uncertainty and $\widetilde{\Delta v_r/\sigma_{vr}}$ the median of the same quantity. The standard deviation comes out to about $1.03$, which means that our reported uncertainties are typically accurate to a few percent. This is in contrast to the uncalibrated measurements which we also plot in Fig.~\ref{fig:exp_std}, and which show both significant offset and uncertainty underestimation.

\section{Catalogues} 
\label{sec:Catalogue}
We publish two catalogues along with this paper. The Main catalogue is the catalogue we recommend for the general user. It includes only relatively precise measurements with low chance of being erroneous. For completion we also publish the Extended catalogue, which includes all measurements we have obtained. The catalogues are available through an online table at \url{https://doi.org/10.5281/zenodo.10043238}. The columns included are described in Table~\ref{tab:columns}.
\begin{table*}[]
    \centering
    \caption{Description of the fields included in the final published catalogue. The first 5 columns (indicated by the horizontal line) are included in the Main catalogue. The remaining columns only appear in the Extended catalogue.}
    \begin{tabular}{cll}
    \hline
         Column & Heading & Description  \\
         \hline
         \vspace{-3mm}\\
         1 & \texttt{source\_id} & \gaia DR3 \texttt{source\_id}\\
         2 & \texttt{ra} &  \gaia DR3 right ascension\\
         3 & \texttt{dec} & \gaia DR3 declination\\
         4 & \texttt{rv} & Calibrated XP radial velocity\\
         5 & \texttt{rv\_err} & Calibrated XP radial velocity uncertainty\\
         \hline\\
         6 & \texttt{offset} & Offset applied to the original radial velocity measurement during calibration\\
         7 & \texttt{underestimation\_factor} & Factor ($a$) applied to the measured uncertainties according to Eq.~\ref{eq:corr uncertainty}\\
         8 & \texttt{noise\_floor} & Noise floor ($c$) applied to the measured uncertainties according to Eq.~\ref{eq:corr uncertainty}\\
         9 & \texttt{CMD\_outlier\_fraction} & Fraction of stars in colour-mag-extinction(-\logg) range that are considered outliers\\
         10 & \texttt{bad\_measurement} & Probability of a bad measurement based on the Random Forest Classifier\\
         11 & \texttt{warning} & Warning flag to indicate potentially problematic radial velocity measurements\\
         12 & \texttt{teff} & \teff of the best fit model for radial velocity\\
         13 & \texttt{logg} & \logg of the best fit model for radial velocity\\
         14 & \texttt{feh} & \feh of the best fit model for radial velocity\\
         15 & \texttt{reduced\_chi\_squared} & Reduced $\chi^2$ of the best fit model for radial velocity\\
         16 & \texttt{skew} & Skewness of the radial velocity posterior probability distribution\\
    \hline
    \end{tabular}
    \label{tab:columns}
\end{table*}

\subsection{Main catalogue}
\begin{figure}
    \centering
    \includegraphics[width=\linewidth]{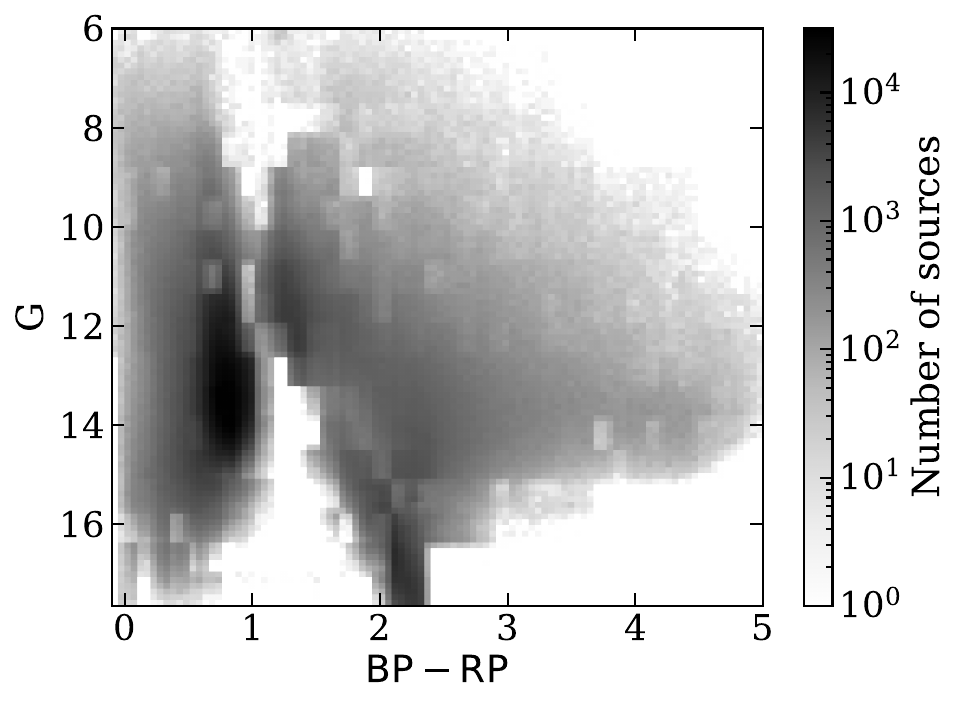}
    \caption{Colour-magnitude diagram for the sources appearing in our Main catalogue.}
    \label{fig:colour-mag_main}
\end{figure}

To ensure we only publish relatively high quality measurements in our Main catalogue, we apply the following selections:
\begin{itemize}
    \item \ebv$ < 0.5$
    \item \texttt{rv\_err} $ < 300$ \kms
    \item \texttt{CMD\_outlier\_fraction} $<0.2$
    \item \texttt{bad\_measurement} $<0.1$.
\end{itemize}
The Main catalogue contains 6\,367\,355 sources that pass the quality cuts. About 23\% of these sources have no previous measurement in \gaia RVS, by far the biggest catalogue in our magnitude range. This means the Main catalogue contains relatively accurate and precise radial velocity measurements for about $1.5$M sources that have no previous measurement available. In Fig.~\ref{fig:colour-mag_main} we show the colour-magnitude density of our Main catalogue.

\subsection{Extended catalogue}
\begin{figure}
    \centering
    \includegraphics[width=\linewidth]{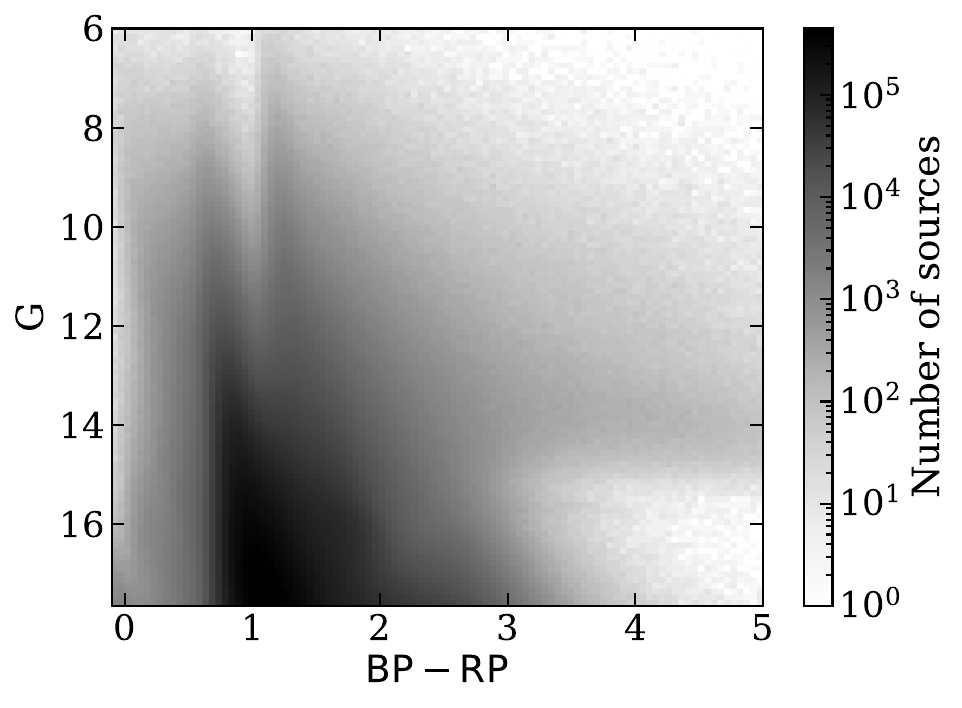}
    \caption{Colour-magnitude diagram for the sources appearing in the Extended catalogue.}
    \label{fig:colour-mag_ext}
\end{figure}

For completion we also provide our entire catalogue, without any quality cuts, which we refer to as our Extended catalogue. The only exception is that we still only publish sources with \ebv$<0.5$, since we deem most higher extinction measurements to be unreliable. To assist the user in making use of this catalogue, we provide additional parameters for all sources alongside those provided for the main catalogue, which is a subset of the Extended catalogue.
In the case of a star occupying a point in parameter space with insufficient reference measurements to perform calibration, we still report our XP radial velocity measurement, only without any calibration performed. In those cases we report the \texttt{CMD\_outlier\_fraction}, \texttt{offset}, \texttt{underestimation\_factor}, and \texttt{noise\_floor} parameters as NaN values.
In Fig.~\ref{fig:colour-mag_ext} we show the colour-magnitude diagram of the sources appearing in our Extended catalogue.
Since there are many more caveats with this data set compared to the Main catalogue, we provide the user with the \texttt{warning} parameter that is provided as a bitmask. If one of the following conditions is met, the corresponding bit is set to 1.
\begin{enumerate}
    \item No calibration applied (0001)
    \item Neighbour in \gaia within 2 arcsec (0010)
    \item \texttt{CMD\_outlier\_fraction} $ > 0.2$ (0100)
    \item \texttt{bad\_measurement} $ > 0.1$ (1000)
\end{enumerate}

\section{Finding hypervelocity stars}
\label{sec:HVS}
Having presented our Main catalogue, we will now apply it to the science case of hypervelocity stars (HVSs).\

HVSs can travel with velocities well in excess of $1000$ \kms \citep{Koposov_2020}, making them much faster than stars belonging to other populations. These stars are ejected from the Galactic Centre following a dynamical encounter with our central massive black hole, Sgr A* \citep[][]{Brown_2015}. Their identification has proven difficult with only a few dozen promising candidates \citep{Brown_2014} and a single star that can be unambiguously traced back to the centre of our Galaxy \citep{Koposov_2020}. Our new catalogue of radial velocities can facilitate blind searches for additional HVSs, helping to unravel the dynamics and properties of stars in the centre of our Galaxy as well as providing valuable information about the Galactic potential \citep[e.g.,][]{Rossi_2017, Evans_2022}.\ 

For the purpose of searching for HVSs it is of interest to determine if we can still obtain reliable radial velocity measurements for extremely high velocity stars. To date, \object{S5-HVS1} is the fastest unbound star known in our Galaxy, with a total velocity in the Galactic frame of $1755\pm50$ \kms and a heliocentric radial velocity of $1017\pm2.7$ \kms \citep{Koposov_2020}. The calibrated radial velocity we measure is $799\pm 273$ \kms, which is consistent with the reference radial velocity measurement of \object{S5-HVS1} within $\sim0.8\sigma$. The \texttt{bad\_measurement} parameter from the RFC is $0.02$ for \object{S5-HVS1}, indicating a reliable measurement. This establishes that our results are still accurate for extremely high radial velocity sources.\

\begin{figure}
    \centering
    \includegraphics[width=\linewidth]{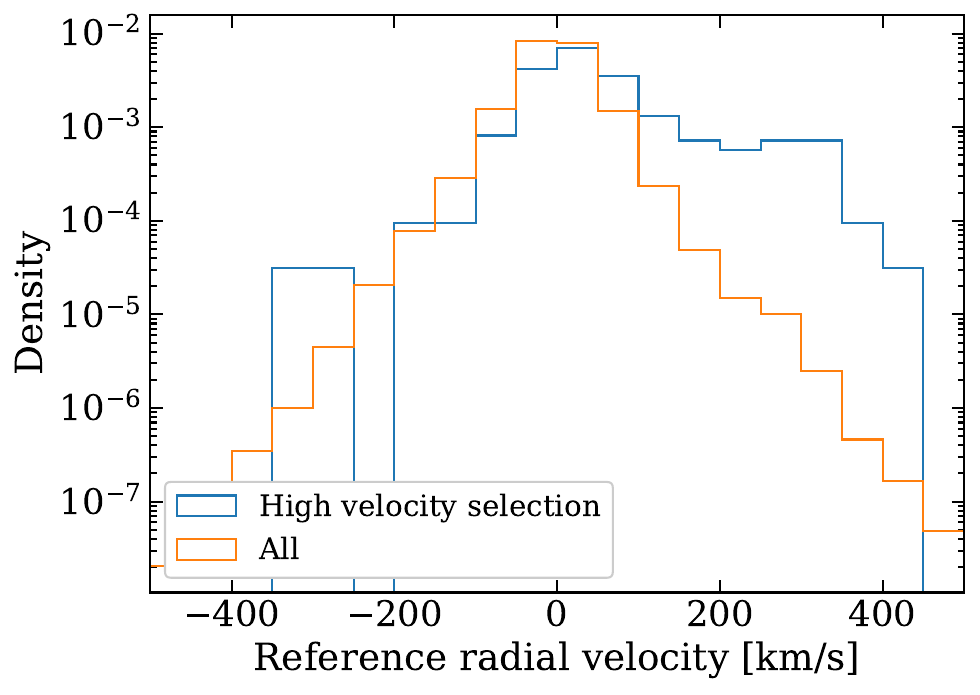}
    \caption{Density of the reference radial velocity distribution of all sources against those selected by $v_\textrm{r}>300+3\cdot\sigma_\textrm{vr}$ \kms in our Main catalogue and for which reference radial velocity measurements are available.}
    \label{fig:high_vel_selection}
\end{figure}

To evaluate the general effectiveness of the selection of high radial velocity star candidates from our Main catalogue, we produce Fig.~\ref{fig:high_vel_selection}.
In the figure we only select those stars from our Main catalogue whose $3\sigma$ lower limit on $v_\textrm{r}$ is at least 300 \kms, i.e., $v_\textrm{r}>300+3\cdot\sigma_\textrm{vr}$ \kms and plot those for which reference radial velocity measurements are available. Although the distribution of our selection still peaks at 0 \kms, we can see a very significant over-density of high radial velocity sources. Most of the sources in this selection do not have reference radial velocity measurements and, as Fig.~\ref{fig:high_vel_selection} shows, the majority of them will not have high radial velocities. However, the selection of HVS candidates for follow-up radial velocity surveys can be viable. Only 3175 sources out of our Main catalogue of 6.4M sources pass the selection of $v_\textrm{r}>300+3\cdot\sigma_\textrm{vr}$ \kms. The number of candidates could be further reduced by using e.g.~astrometric information to constrain the orbits. Follow-up observations will be proposed for the most promising of these HVS candidates to precisely measure their radial velocities.

\section{Discussion}
\label{sec:Discussion}
Despite the challenges, we have shown that radial velocities can be obtained from \gaia XP measurements to a precision of better than $\sim300$ \kms for stars as faint as $\textrm{G}=17.65$. Sect.~\ref{sec:improvements_of_the_method} discusses possible improvements to the methods presented in this paper, in Sect.~\ref{sec:dr4} we provide prospects for \gaia DR4 and the improvements that we might expect with its release regarding radial velocities from XP spectra, and lastly in Sect.~\ref{sec:science_cases} we discuss science cases for XP radial velocities in \gaia DR4.

\subsection{Improvements to the method}
\label{sec:improvements_of_the_method}
Our current approach is only viable for low to intermediate extinction sources, due to our implementation of an initial guess on \teff. This approximation breaks down for high extinction sources as mentioned in Sect.~\ref{sec:spectral_analysis}. Practically, this means that our results are not reliable for sources with \ebv $\gtrsim 0.5$ and we do not report our results for those sources. The issue can be mitigated by e.g.,~using a larger range in \teff during the fitting procedure for high extinction sources, or by fitting every model for all sources. 

Additionally, fitting for extinction rather than relying on a 2D extinction map would allow for more accurate measurements, because for individual sources the 2D extinction map is only an estimate of the actual line-of-sight extinction. Including extinction in the fitting procedure is possible, but is also computationally very expensive, which is why we opted to use the 2D map instead. 

The analysis could be further improved by choosing the fitting wavelength range on a source-to-source basis: practically one would chose for each source the wavelength regions that hold the most spectral information. Doing this would improve the precision of the radial velocity measurements. Here, we used instead the same wavelength ranges throughout.

Alternatively to the modelling presented in this work, one could forward model the spectral coefficients directly. Provided that the design matrix and the model of the instrument are accurate, this should give more precise results.

Improving upon the method is required if the goal is to obtain reliable radial velocity measurements for a revolutionary large set of sources. Even though we started out with XP spectra to about 220M sources, the Main catalogue only includes around $6.4$M radial velocities (or about $3\%$ of XP sources), with an additional $\sim119$M in the Extended catalogue. Understanding and correcting for systematic effects remains the most challenging aspect.

\subsection{\gaia Data Release 4}
\label{sec:dr4}
In \gaia DR4, XP spectra will be published for about 2B sources, in addition to individual epoch spectra of said sources. Since this is orders of magnitude higher than any current radial velocity catalogue, the potential scientific return on a well optimised method of radial velocity analysis would be very high.

It is unknown how large the improvement will be in radial velocity accuracy and precision from \gaia DR3 to DR4 using the methods presented here. The reason is that systematics are a very large factor in the radial velocity uncertainty. We might consider the noise floor in Eq.~\ref{eq:corr uncertainty} to be the intrinsic systematic uncertainty caused by imperfect calibration of XP spectra in \gaia DR3. If we assume that these systematics are resolved in \gaia DR4 we can provide an outlook for the performance of our method when applied to \gaia DR4. When we ignore the noise floor, the number of sources with radial velocity uncertainties $<300$ \kms in DR3 approximately doubles to $\sim17$M. In addition, there would be about 1M sources with radial velocity uncertainties $<100$ \kms. The smallest uncertainties that might be achieved are expected to be on the order of 50 \kms. Although DR4 will mostly include fainter sources than the current limit of $G<17.65$, the S/N for a given magnitude will also improve. \gaia DR4 will provide XP spectra to about 9 times as many sources as DR3. A rough approximation of the final number of sources with a particular quality in \gaia DR4 is thus 9 times the number in DR3. This would imply that the XP spectra in \gaia DR4 could provide us with $\sim153$M and $\sim8$M measurements with uncertainties better and 300 and 100 \kms respectively. Without systematic uncertainty due to the noise floor, these stars would be mainly red (\bprp $\gtrsim2$) and blue (\bprp $\lesssim0.7$) in colour. 

\subsection{Science cases in \gaia Data Release 4}
\label{sec:science_cases}
Having discussed improvements to both the methods and data with the next data release of \gaia, we will now look at prospects for two specific science cases in \gaia DR4: dark companions and HVSs.\

Dark companions refer to binary systems in which one of the components emits little to no light in the photometric band used to observe them. These dark companions, such as black holes, can be identified from low-resolution spectra if enough epochs are available over a sufficient time span. The photocentre of \gaia \object{BH1}, for instance, has a radial velocity amplitude of about $130$ \kms \citep{El-Badry_2023, Chakrabarti_2023}, far larger than the typical uncertainty in \gaia RVS of only a few \kms \citep{Katz_2023}. With the release of epoch XP spectra in \gaia DR4, searches for dark companions will become possible in the full \gaia catalogue of $\sim2$B sources. Compared to the astrometric time-series, radial velocities have the advantage of being distance independent, thus allowing for a larger search volume. Also, in comparison to \gaia RVS, the XP radial velocities have the advantage of being deeper and therefore covering a larger volume. \gaia RVS will have a limiting magnitude of $\textrm{G}_\textrm{RVS}\sim16$ in \gaia DR4, compared to the limiting magnitude of $\textrm{G}\sim20.7$ for the XP spectra. We assume the two photometric bands to be similar\footnote{\url{https://www.cosmos.esa.int/web/gaia/dr3-passbands}} and approximate the magnitude difference as $4.7$. From the magnitude difference we can calculate the volume ratio as 
\begin{equation}
\frac{V_\textrm{XP}}{V_\textrm{RVS}} = 10^{0.6\cdot\Delta m} \approx 660,
\end{equation}
with $V_\textrm{XP}$ the volume covered by XP spectra, $V_\textrm{RVS}$ the volume covered by RVS radial velocities, and $\Delta m$ the difference in limiting magnitude. The effective volume covered by XP spectra is thus about 660 times as large as that covered by RVS radial velocities. Depending on the final precision and accuracy that can be achieved, dedicated higher resolution observations might be required to confirm systems with possible dark companions identified from \gaia XP radial velocities.\

In addition to finding dark companions, \gaia DR4 XP radial velocities could support the search for HVSs. Because of the high intrinsic velocities of these stars, large uncertainties are less problematic. As demonstrated in Sect.~\ref{sec:HVS}, the contamination of a selection of extremely high XP radial velocity sources is substantial in our Main catalogue. With an improved analysis as suggested in Sect.~\ref{sec:improvements_of_the_method}, in combination with a reduction in systematics that we expect in \gaia DR4, the contamination will decrease. This will allow for more effective follow-up campaigns to identify new HVSs. As mentioned above, the advantage of using \gaia XP spectra, is that the effective volume is much larger than \gaia RVS.\

Both for dark companions and HVSs, XP radial velocities will be most effective in identifying them for red (\bprp $\gtrsim2$) and blue (\bprp $\lesssim0.7$) sources, since these sources have the lowest uncertainties in XP radial velocity. This is not expected to change from \gaia DR3 to DR4, since it is inherent to the radial velocity information contained within the XP spectra.

\section{Conclusion}
\label{sec:Conclusion}
As a proof of concept, we have clearly demonstrated that \gaia XP spectra can be used to measure radial velocities. In this work we publish the Main catalogue containing reliable and precise radial velocity measurements for about $6.4$M sources, 23\% of which have no previous radial velocity measurements in \gaia. In addition, we publish the Extended catalogue containing all $\sim125$M sources for which we have obtained a radial velocity measurement. This constitutes $\sim84$\% of sources with \gaia XP spectra and \ebv $<0.5$. The extended catalogue, however, contains a significant number of unreliable measurements and should therefore only be used with caution.

In general, sources with \bprp $\gtrsim2$ and \bprp $\lesssim0.7$ tend to give the most precise radial velocity measurements in our catalogue, down to uncertainties $\sim100$ \kms. In the future, we expect the most precise radial velocity measurements from \gaia XP spectra to have uncertainties on the order of 50 \kms.

Critically, this work has demonstrated the potential of measuring radial velocities for over $10^9$ sources in \gaia DR4 using XP spectra. This would constitute an orders of magnitude increase compared to the largest current catalogue. Before then, the methods presented here should be further improved to fully exploit the scientific content available to us.

\begin{acknowledgements}
The authors thank the anonymous referee for their insightful comments and suggestions on this work. In addition, the authors would like to thank the attendees at the \gaia XPloration workshop for their input and enthusiasm. Special thanks goes to Francesca De Angeli, Anthony Brown, and Vasily Belokurov for their support, helpful insight, and discussions. We would also like to thank Anthony Brown for his feedback on a first draft of this manuscript.\\

EMR acknowledges support from European Research Council (ERC) grant number: 101002511/project acronym: VEGA\_P. TM acknowledges a European Southern Observatory (ESO) fellowship.
This work has made use of data from the European Space Agency (ESA) mission
{\it Gaia} (\url{https://www.cosmos.esa.int/gaia}), processed by the {\it Gaia}
Data Processing and Analysis Consortium (DPAC,
\url{https://www.cosmos.esa.int/web/gaia/dpac/consortium}). Funding for the DPAC
has been provided by national institutions, in particular the institutions
participating in the {\it Gaia} Multilateral Agreement.
This project was developed in part at the 2023 \gaia XPloration, hosted by the Institute of Astronomy, Cambridge University. This paper made use of the Whole Sky Database (wsdb) created and maintained by Sergey Koposov at the Institute of Astronomy, Cambridge with financial support from the Science \& Technology Facilities Council (STFC) and the European Research Council (ERC). This work was performed using the ALICE compute resources provided by Leiden University. This research or product makes use of public auxiliary data provided by ESA/Gaia/DPAC/CU5 and prepared by Carine Babusiaux. For the purpose of open access, the author has applied a Creative
Commons Attribution (CC BY) licence to any Author Accepted Manuscript version
arising from this submission.\\

Software: \texttt{NumPy} \citep{Harris_2020}, \texttt{SciPy} \citep{2020SciPy-NMeth}, \texttt{Matplotlib} \citep{Hunter_2007}, \texttt{Astropy} \citep{astropy:2013, astropy:2018, astropy:2022}, \texttt{emcee} \citep{Foreman_2013}, \texttt{Numba} \citep{Lam_2015}, \texttt{dustmaps} \citep{Green_2018}, \texttt{GaiaXPy}, \texttt{SpectRes} \citep{Carnall_2017}, \texttt{extinction}\footnote{\url{https://extinction.readthedocs.io/en/latest/}}, \texttt{healpy} \citep{Gorski_2005, Zonca_2019}, \texttt{corner} \citep{corner}, \texttt{scikit-learn} \citep{scikit-learn}.
\end{acknowledgements}

\bibliographystyle{aa}
\bibliography{mybib}

\begin{appendix}

\section{\teff initial guess}
\label{app:teff}
For computational efficiency, we make an initial guess on \teff for each source and only consider models within 500K of this initial guess as described in Sect.~\ref{sec:spectral_analysis}.

We make the initial guess on \teff\ based on the \bprp colour and 2D extinction along the line of sight. First we correct the colour of the source for extinction using the extinction law provided by \gaia\footnote{\url{https://www.cosmos.esa.int/web/gaia/edr3-extinction-law}}. This is only an approximation, since we do not know the intrinsic colour of a source affected by extinction and use the observed colour instead. By analysing test sources for the entire \teff\ grid range, we used an empirical exponential fit to the relation between \bprp and \teff. Only for colours \bprp $\lesssim0$ we found a turn-off from the exponential relation, for which we fit a linear function. The relation is given by
\begin{equation}
 \teff(K) = 
  \begin{cases} 
   6721\cdot\exp{\left(-0.95\cdot c\right)} + 2617 & \text{if } c > 0.01 \\
   -18796\cdot c + 9431       & \text{if } c \leq 0.01
  \end{cases}
\end{equation}
where $c$ denotes the extinction corrected \bprp colour. If the initial guess for \teff\ is outside the temperature range of our models (see Table~\ref{tab:grid edges}) we do not attempt to fit the source and no radial velocity is obtained. 

\section{\logg systematics}
\label{app:logg_systematics}
\begin{figure}[h!]
\centering
\includegraphics[width=\linewidth]{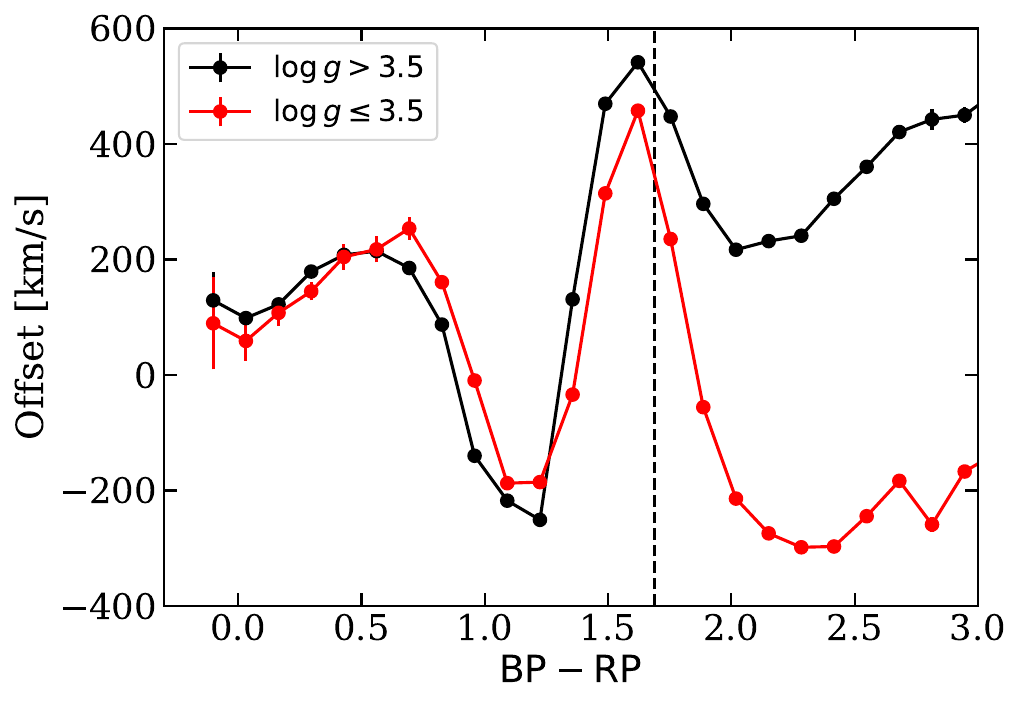}
\caption{The offset between the uncalibrated \gaia XP radial velocities and reference ones as a function of \bprp colour. We split the dwarfs and the giants using the \logg measurements from \citet{Zhang_2023}. The dashed vertical line shows the limit above which we use 2 bins in \logg for our calibration model. The figure only includes sources with \ebv$<0.1$ and $12 < \textrm{G} < 13$.}
\label{fig:split_colour}
\end{figure}

\begin{figure}
\centering
\includegraphics[width=\linewidth]{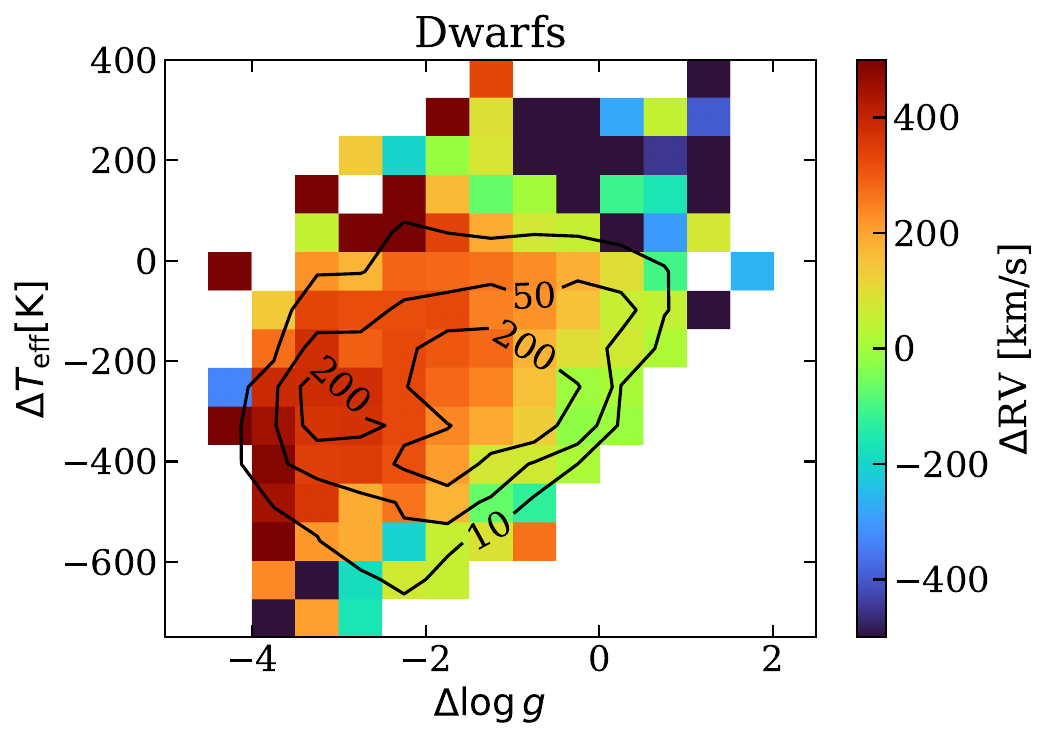}
\caption{Here we show the radial velocity offset for dwarfs (\logg$>3.5$ according to \citet{Zhang_2023}) as a function of the difference in the best-fit \logg and \teff we obtain versus those from APOGEE. The contour lines give the underlying source density. The stars in the sample have $10<\textrm{G}<16$, $2<$ \bprp $<2.5$, and \ebv$<0.1$.}
\label{fig:apogee_diff_dwarfs}
\end{figure}

For red sources we observe strong systematics in the radial velocity we measure, based on the surface gravity of the stars. We demonstrate this in Fig.~\ref{fig:split_colour}, where we use the surface gravity measurements from \citet{Zhang_2023} to separate dwarfs from giants.
To investigate the origin of this systematic offset we used \logg and \teff measurements from APOGEE in comparison to our best-fit values. In Fig.~\ref{fig:apogee_diff_dwarfs} we show how the radial velocity offset relates to errors in the parameter estimation for dwarfs.
We can see that dwarfs tend to get assigned radial velocities that are too high. This also varies as a function of $\Delta$\logg and $\Delta$\teff. In particular, if the \logg of our best fit model labels a dwarf as a giant a positive offset is introduced, which can be seen from the colour gradient towards negative $\Delta$\logg.

\begin{figure}
    \centering
    \includegraphics[width=\linewidth]{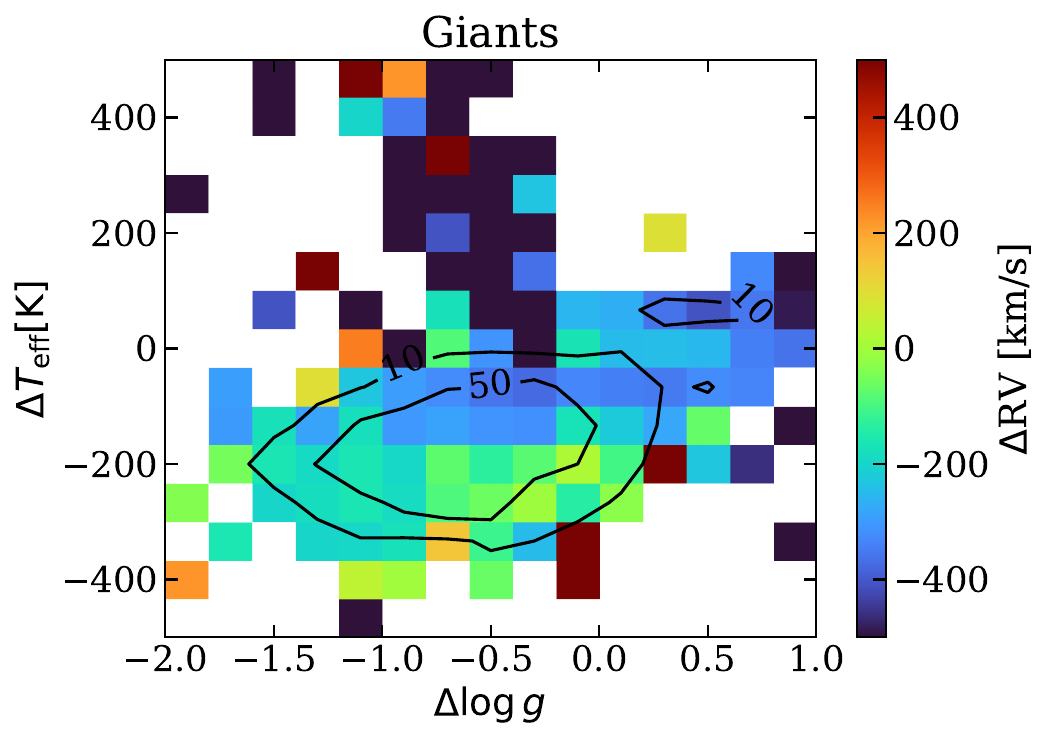}
    \caption{Same as Fig.~\ref{fig:apogee_diff_dwarfs}, but now only for giants.}
    \label{fig:apogee_diff_giants}
\end{figure}

\begin{figure}
    \centering
    \includegraphics[width=\linewidth]{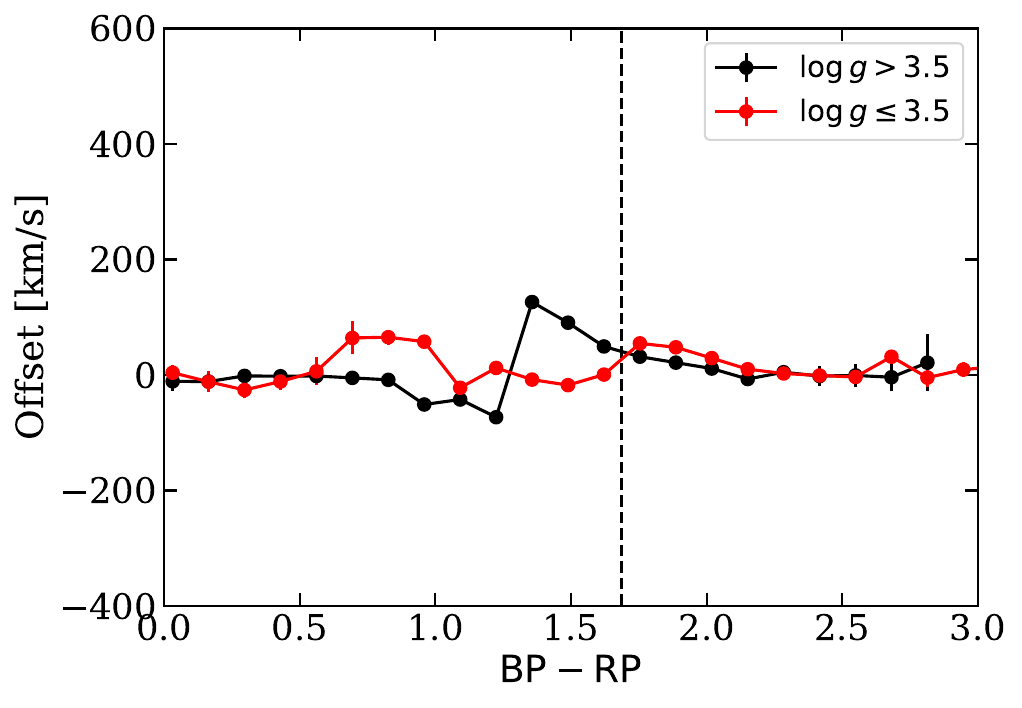}
    \caption{The same as Fig.~\ref{fig:split_colour}, but now after calibration has been applied for the bin $11.9575<\textrm{G}<12.59$.}
    \label{fig:split_colour_corr}
\end{figure}

To make this figure and have sufficient sources, we used a much larger range of colour and magnitude than we do for single bins in the calibration. The spread in offsets for individual bins is smaller and therefore less problematic. Fig.~\ref{fig:apogee_diff_giants} shows the same, but now only for giants.
In general we can see that the offset tends to be much lower for giants. It is possible that it is caused by the basis-function representation in \gaia, which undergoes optimisation and might lead to systematic differences in the translation of giant and dwarf spectra. Fortunately, we can effectively mitigate the effects of this bias, whatever its origin. To evaluate our treatment of the observed offset described in Sect.~\ref{sec:calibration_models} we recreate Fig.~\ref{fig:split_colour} for our calibrated sample, which we show in Fig.~\ref{fig:split_colour_corr}.
The figure clearly demonstrates that the offset between dwarfs and giants for very red sources has been effectively removed. Colour dependent systematics are still visible in the offset, however, particularly around \bprp$\sim1.5$. This effect is explained in Sect.~\ref{app:MCMC} and is related to the \texttt{CMD\_outlier\_fraction}.

\section{Further validation}
\begin{figure}[h!]
    \centering
   \includegraphics[width=\linewidth]{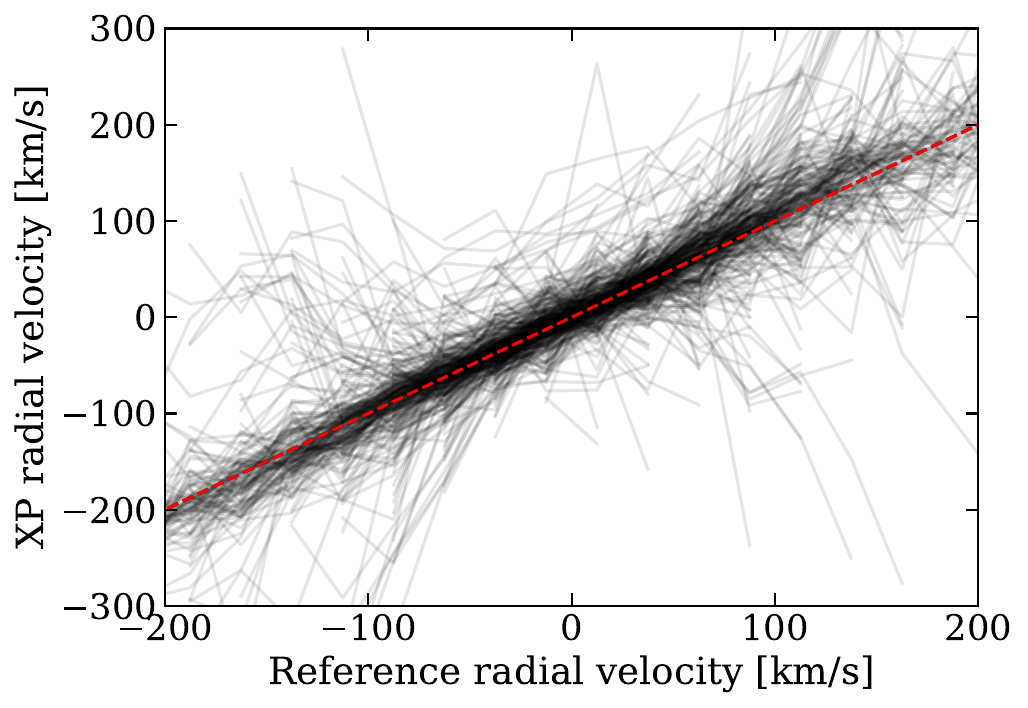}
    \caption{The median XP radial velocity as a function of bins in reference radial velocity. The dashed red line gives the bisection and the transparent black lines the data for individual bins in the calibration.}
    \label{fig:veri_rv_sens}
\end{figure}

To demonstrate that the calibration we perform is not introducing an apparent radial velocity sensitivity, we include Fig.~\ref{fig:veri_rv_sens}.
This figure gives the median XP radial velocity for bins in reference radial velocity within the Main catalogue, where each curve is constructed using stars within one colour-magnitude diagram bin. The one-to-one slope demonstrates that within individual colour-magnitude bins we are sensitive to radial velocity. This excludes the possibility that a correlation between colour-magnitude and radial velocity is introducing an apparent radial velocity sensitivity.\\
\FloatBarrier

\section{Markov chain Monte Carlo examples and results}
\label{app:MCMC}
\begin{figure*}
\centering
\includegraphics[width=\linewidth]{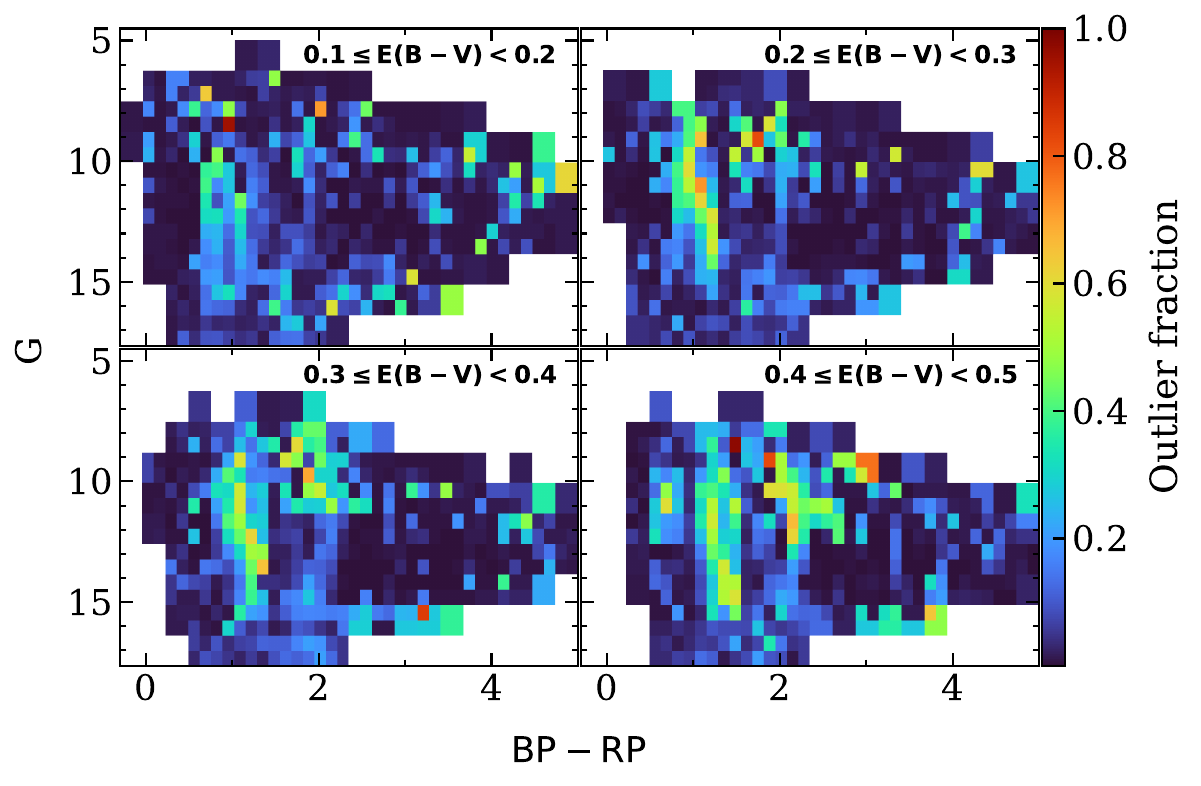}
\caption{The outlier fractions as a function of colour and magnitude for the remaining \ebv bins not shown in the main text.}
\label{fig:outliers_app}
\end{figure*}

In Fig.~\ref{fig:outliers_app} we show the outlier fraction of the model described in Sect.~\ref{sec:calibration_models}, Eq.~\ref{eq:likelihood} as a function of colour and magnitude for the higher \ebv bins, not shown in the main text.

\begin{figure*}
\centering
\includegraphics[width=\linewidth]{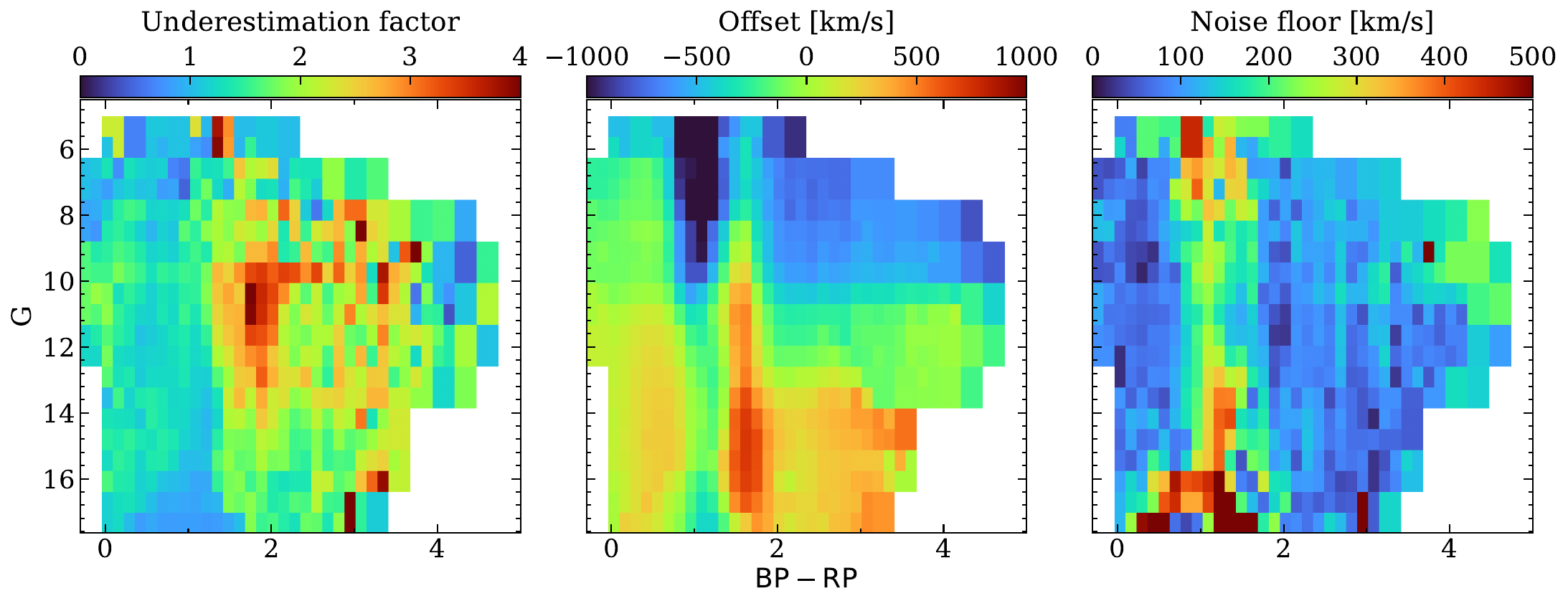}
\caption{From left to right, the underestimation factor, offset, and noise floor for sources with \ebv$<0.1$.}
\label{fig:2d_params}
\end{figure*}

In Fig.~\ref{fig:2d_params} we give the underestimation factor, offset, and noise floor for sources with \ebv$<0.1$. The definition of the offset can be found in equation~\ref{eq:likelihood} and the definition of the underestimation factor and noise floor in equation~\ref{eq:corr uncertainty}.

\begin{figure*}
\centering
\includegraphics[width=\textwidth]{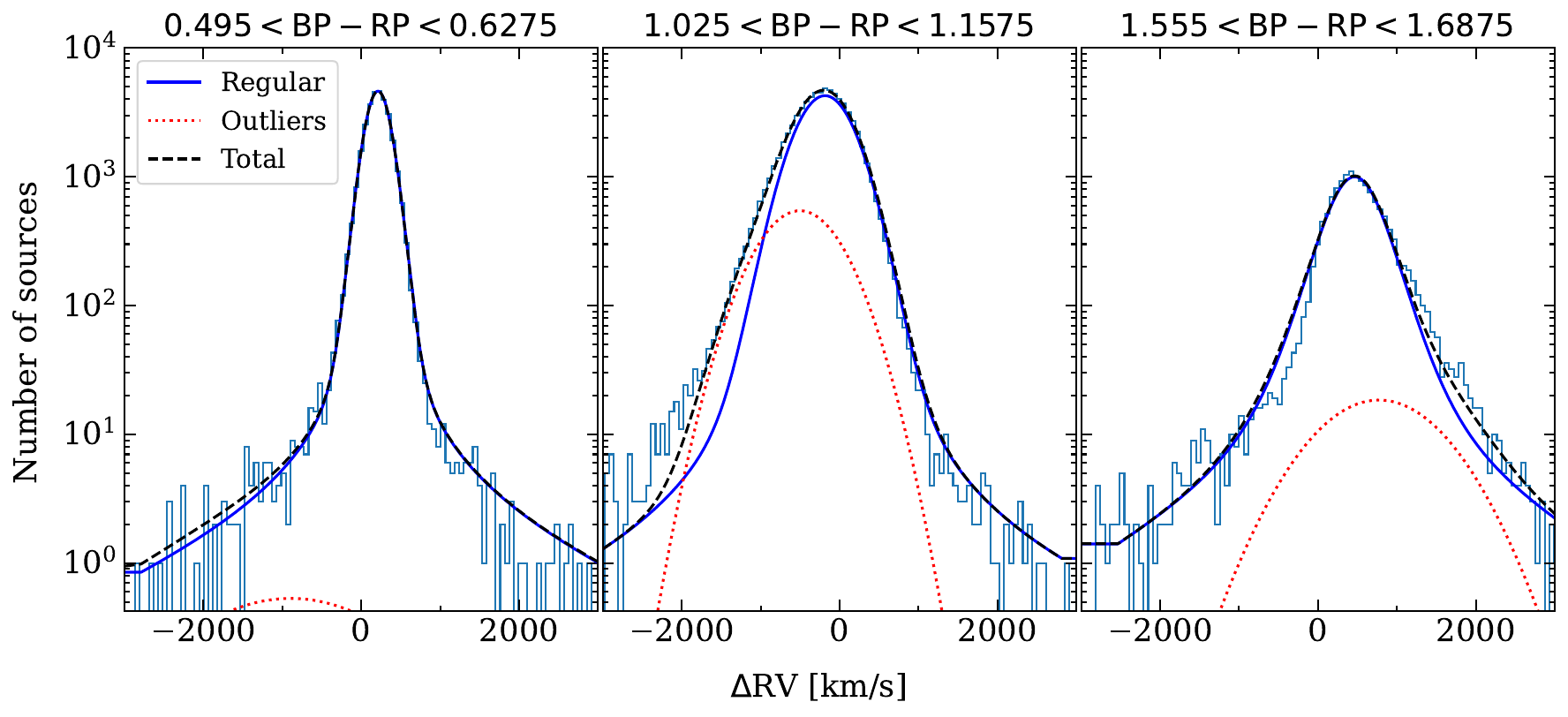}
\caption{Histograms of the radial velocity difference between the raw XP measurement and the reference values. The three lines show the parameters found by the fit of the predicted distributions of the regular, outlier, and combined populations. The different panels show different bins in \bprp colour, as indicated above the panels.}
\label{fig:MCMC_examp}
\end{figure*}

To create more insight into what our calibration model does, we show here the MCMC results for a number of bins. We choose to show the $11.9575<\textrm{G}<12.59$ magnitude bin for three different colour bins indicated in Fig.~\ref{fig:MCMC_examp}. 
Not for all bins does the method work equally well, which affects the calibration. If we look at the middle and right panels, we see that the overall distribution is skewed. Correcting for the offset in that sample will still result in a median that is significantly higher or lower than 0, because of the skew. This is the residual effect we see in Fig.~\ref{fig:split_colour_corr}. Part of this can be explained by \logg effects that are still present in this colour region. We chose not to attempt to correct for these effects, because for these intermediate colour sources, there is no clear split in offset between dwarfs and giants. Instead a more elaborate strategy would have to be employed to correct for the remaining systematics.

\end{appendix}
\end{document}